\documentclass[conference, a4paper]{IEEEtran}
\IEEEoverridecommandlockouts

\usepackage{cite}
\usepackage{eso-pic}
\usepackage{amsmath,amssymb,amsfonts,amsthm}
\usepackage{orcidlink}
\usepackage[normalem]{ulem}
\usepackage{textgreek}
\usepackage{upgreek}
\usepackage[linesnumbered,ruled,vlined]{algorithm2e}

\SetNlSty{}{}{}
\SetKwInput{KwInput}{Input}
\SetKwInput{KwOutput}{Output}
\SetKwComment{Comment}{$\triangleright$\ }{}

\SetCommentSty{fontcommand}
\newcommand{\negVeryTiny}{\kern -0.5pt}

\newlength{\PZEROsubjectToSpaceOptprob}
\newlength{\PZEROconstraintsSpaceOptprob}
\newlength{\subjectToSpaceOptprob}
\newlength{\addLengthSpaceOptProb}
\newlength{\addLengthSpaceOptProbNeg}
\newlength{\hphantomSt}
\newlength{\constraintsSpaceOptprob}
\newlength{\constraintsSpaceSmallerBracketsOptprob}
\usepackage[dvipsnames]{xcolor}
\usepackage{tikz}
\usetikzlibrary{positioning, arrows.meta}
\usepackage{contour}
\usepackage{pgfplots}
\usepackage[position=top]{subfig}
\usepackage{graphicx}
\usepackage[printonlyused,nohyperlinks,nolist]{acronym}
\usepackage{textcomp}

\usepackage{siunitx}
\usepackage{booktabs}
\usepackage{multirow}
\usepackage{multicol}
\usepackage{enumitem}
\usepackage{tabularx}
\usepackage{bm}
\usepackage{mathtools}

\newcolumntype{L}[1]{>{\raggedright\arraybackslash}p{#1}}
\newcolumntype{C}[1]{>{\centering\arraybackslash}p{#1}}
\newcolumntype{R}[1]{>{\raggedleft\arraybackslash}p{#1}}

\newlist{mylist}{itemize}{1}
\setlist[mylist]{label=\textbullet,
                 nosep, wide, leftmargin=*,
                 before=\vspace{-0.57\baselineskip},
                 after=\vspace{-0.85\baselineskip}}

\def\BibTeX{{\rm B\kern-.05em{\sc i\kern-.025em b}\kern-.08em
    T\kern-.1667em\lower.7ex\hbox{E}\kern-.125emX}}

\usepackage{hyperref}
\hypersetup{
    colorlinks,
    linkcolor={red!50!black},
    citecolor={blue!50!black},
    urlcolor={blue!80!black},
    draft
}

\newcommand{\figref}[1]{Fig.~\ref{#1}}
\newcommand{\secref}[1]{Section~\ref{#1}}

\newcommand{\tabref}[1]{Table~\ref{#1}}

\newcommand{\set}[1]{{\mathcal{#1}}}

\makeatletter
\newcommand{\subalign}[1]{%
  \vcenter{%
    \Let@ \restore@math@cr \default@tag
    \baselineskip\fontdimen10 \scriptfont\tw@
    \advance\baselineskip\fontdimen12 \scriptfont\tw@
    \lineskip\thr@@\fontdimen8 \scriptfont\thr@@
    \lineskiplimit\lineskip
    \ialign{\hfil$\m@th\scriptstyle##$&$\m@th\scriptstyle{}##$\hfil\crcr
      #1\crcr
    }%
  }%
}
\makeatother

\AddToShipoutPictureFG{%
  \AtPageUpperLeft{%
    \raisebox{-7mm}{%
      \makebox[\paperwidth][c]{%
        \parbox{\dimexpr\paperwidth-4cm\relax}{%
          \centering
          \color{gray}
          \fontsize{7}{7}\selectfont
          This work has been submitted to the IEEE for possible publication. Copyright may be transferred without notice, after which this version may no longer be accessible. 
        }%
      }%
    }%
  }%
}

\IEEEaftertitletext{\vspace{-1.5em}}

\begin{document}
\bstctlcite{IEEEexample:BSTcontrol}
\IEEEaftertitletext{\vspace{-1\baselineskip}}

\title{Multi-User Linear Precoder Optimization for Obfuscation-Aided Mixed Regular and Wiretap Coding}

\author{\IEEEauthorblockN{Bastian Loss, Ahmad Nimr, and Gerhard Fettweis}
\IEEEauthorblockA{Vodafone Chair Mobile Communications Systems, Technische Universit{\"a}t Dresden, Germany\\
Email: \{bastian.loss,  ahmad.nimr, gerhard.fettweis\}@tu-dresden.de}
}

\vspace{-2em}

\maketitle

\begin{acronym}
    \acro{MISO}{multiple-input single-output}
    \acro{SIC}{successive interference cancellation}
    \acro{BC}{broadcast channel}
    \acro{SCA}{successive convex approximation}
    \acro{BS}{base station}
    \acro{CSI}{channel state information}
    \acro{PLS}{physical layer security}
    \acro{DL}{downlink}
    \acro{MUI}{multi-user interference}
    \acro{MA}{multiple access}
    \acro{OMA}{orthogonal multiple access}
    \acro{RSMA}{rate-splitting multiple access}    
    \acro{SINR}{signal-to-interference-plus-noise ratio}
    \acroplural{SINR}[SINRs]{signal-to-interference-plus-noise ratios}
    \acro{SNR}{signal-to-noise ratio}
    \acroplural{SNR}[SNRs]{signal-to-noise ratios}
    \acro{AE}{authenticated encryption}
    \acro{AWGN}{additive white Gaussian noise}
    \acro{MRT}{maximum ratio transmission}
    \acro{ZF}{zero forcing}
    \acro{MU}{multi-user}
    \acro{LP}{linear precoding}
    \acro{WT}{wiretap}
    \acro{SC}{superposition coding}
    \acro{SRC}{successive refinement coding}  
    \acro{SI}{secret information}  
    \acro{LoS}{line-of-sight}  
    \acro{UPW}{uniform plane-wave}
    \acro{ULA}{uniform linear array}
    \acro{CCP}{convex concave procedure}
    \acro{OTP}{one-time-pad}
    \acro{QoS}{quality of service}
\end{acronym}
\begin{abstract}\label{abstract}
Physical-layer security has attracted considerable research interest with the development of quantum computing. One prominent approach is wiretap coding, which exploits the channel advantage of a legitimate receiver over an eavesdropper to achieve secure communication. However, under unfavorable channel conditions, wiretap coding can significantly reduce the achievable rate compared to regular channel coding. Mixed coding alleviates this limitation by using wiretap coding to protect a short secret required for obfuscation, while the obfuscated payload is encoded using regular channel coding.
In this work, we propose a generalized obfuscation-aided mixed-coding system based on the \ac{MU}-\ac{LP} architecture that unifies existing approaches employing different obfuscation functions. We then formulate the corresponding precoder optimization problem, abstracting the obfuscation function through a minimum secrecy rate constraint.
Numerical simulations using a geometric two-user \ac{LoS} channel demonstrated how the minimum secrecy rate requirement determines the operating point of mixed coding between regular and wiretap coding. Subsequent beampattern analysis revealed a key synergy between mixed coding and precoding: Mixed coding enables individual beam shaping for the two coding layers. Finally, spatial analysis showed that mixed coding offers the largest sum-rate gains when users are positioned at similar angles.
\end{abstract}
\begin{IEEEkeywords}
mixed coding, multi-user (MU), linear precoding (LP), geometric channel,  beampattern.
\end{IEEEkeywords} 
\vspace{-0.8em}
\acresetall
\section{Introduction}
\IEEEPARstart{N}{etwork} traffic has become increasingly heterogeneous over the past decades due to the ongoing digitization. In recent years, this development has steadily expanded to include sensitive information, such as health records and authentication credentials. As this trend continues, ensuring the security of network traffic becomes ever more important. For decades, cryptography has provided the foundation for this security. However, advances in quantum computing pose a growing threat to existing cryptographic systems, particularly those relying on asymmetric cryptography \cite{nistsp800175BRev1}. 

These concerns have led to increased interest in physical layer security, which can be used to replace cryptography in various tasks, such as providing confidentiality, providing authentication, or generating keys \cite{Duan2025}. In this work, we focus on the mechanism of \ac{WT} coding, which can provide confidentiality against an eavesdropper with worse channel conditions than the intended user. Wyner showed in his seminal work \cite{Wyner1975} that, under these conditions, a stochastic wiretap encoder can utilize random dummy bits to overload the degraded eavesdropper channel, while the intended user can still reliably decode the message. However, sending dummy bits also reduces the rate available for data transmission. Therefore, relying on \ac{WT} coding can significantly reduce the net data rate of the intended user if the channel conditions are not sufficiently favorable.

To alleviate this problem, an external \emph{obfuscation} mechanism can be integrated into the system that protects parts of the user messages, while other parts are secured through \ac{WT} coding. This idea was developed in several research branches independently using different obfuscation mechanism. \looseness=-1

For example, Tajbakhsh et al. \cite{Tajbakhsh2018} introduced network coding as an obfuscation mechanism. The authors encode the user messages using a network code and divide the resulting linear combinations into two subsets. One subset is protected the information using \ac{WT} coding, while the other layer uses regular channel coding. \ac{WT} then prevents an eavesdropper from obtaining one message subset, preventing decoding of the network code.
 
Layered video coding based on \ac{SRC} was proposed as another obfuscation method by Ng et al. \cite{Ng2016}. This approach decomposes the user message into a base layer and dependent enhancement layers that progressively refine message quality. The authors then propose to only secure the base layer using \ac{WT} coding.
However, existing layered video coding is not designed to provide secrecy of the enhancement layers, which may therefore leak information about the original message. Therefore, this approach should not be considered an adequate obfuscation mechanism. 

Another obfuscation function considered in the literature is cryptographic encryption \cite{Gungor2012, Jeon2025, Bloss2025}. Güngör et al. \cite{Gungor2012} proposed using \ac{OTP} encryption to secure data transmissions when the instantaneous channel conditions are unfavorable. Secret-key bits are generated and buffered during favorable channel conditions and subsequently used to maintain secure communication when channel conditions become unfavorable for the intended user. Jeon et al. \cite{Jeon2025} proposed splitting the user messages into two parts, where one part is protected by \ac{WT} coding, while the other part is protected by encryption. A tradeoff is then considered between power consumption and eavesdropping success probability in satellite systems. \looseness=-1

Our previous work \cite{Bloss2025} employs encryption to enable the use of \ac{RSMA} for secure transmission. In \ac{RSMA}, the data intended for each user is split into a common part and a private part. The common parts are combined into a single common message decoded by all users, while each private part is transmitted separately to its intended user. This structure enables more efficient interference management and can increase the achievable user rate compared with conventional \ac{MU}-\ac{LP} systems that use only private messages \cite{Clerckx2023}. The approach relies on carrying the cryptographic key material in the private messages protected by \ac{WT} coding, while the common message is protected through encryption. Without obfuscation, ensuring the confidentiality of all user messages would prevent the use of the common message, effectively reducing \ac{RSMA} to \ac{WT}-coded \ac{MU}-\ac{LP} \cite{Clerckx2023,Bjoernson2014}.

The overarching conclusion from existing literature is that mixing wiretap and regular coding can increase the achievable rate \cite{Tajbakhsh2018, Bloss2025} or reduce the required transmit power \cite{Ng2016,Jeon2025}. However, to use mixed coding, an obfuscation function is required to secure the data on the regular coding layer. In this work, we incorporate these two ideas into an obfuscation-aided mixed coding system in the \ac{MU}-\ac{LP} architecture. We then formulate the corresponding linear precoder optimization problem, subject to a minimum secrecy rate requirement to ensure that the \ac{SI} required for obfuscation is protected by \ac{WT} coding. From a simulation perspective, we focus on a geometric two-user \ac{LoS} channel. We visualize the resulting beam patterns, as well as spatial user configurations, to investigate how the precoder design amplifies the benefits of mixed coding. This simulations complement existing literature based on statistical channels \cite{Gungor2012,Tajbakhsh2018} or abstract coefficient-based channels \cite{Ng2016, Bjoernson2014, Jeon2025}.

The remaining work is structured as follows.
Section \ref{sec:SystemModel} proposes the general obfuscation-aided \ac{MU}-\ac{LP} system, as well as two special cases.
Section \ref{sec:optproblem} poses the linear precoder optimization problem for all three systems.
Section \ref{sec:ResultsBeampattern} presents the simulation results.
Section \ref{sec:conclusion} concludes this work. 

\begin{figure*}[!t]%
    \centering%
    \captionsetup{justification=centering,labelsep=space}%
    \includegraphics[width=\textwidth]{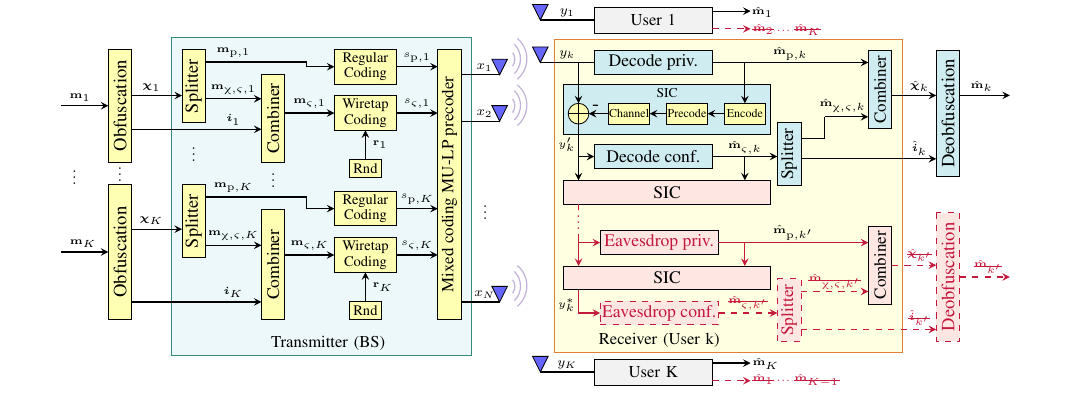}%
    \vspace{-0.5em}%
    \caption{System model for mixed coding \ac{MU}-\ac{LP}}%
    \label{fig:SDMA1HybridFigure}%
    \vspace{-1.5em}%
\end{figure*}%
\section{System model}\label{sec:SystemModel}
For this work we consider a linearly precoded downlink \ac{MISO}-\ac{BC}, which employs a \ac{BS} with $N$ antennas that provides multiple access to $k \in  \mathcal{K} = \{1,\ldots, K\}$ single-antenna users. It is assumed that all user data must be secured without exception. Further, we follow the assumptions of common \ac{MU}-\ac{LP} and \ac{RSMA} literature \cite{Bjoernson2014, Clerckx2023}, regarding the information that transmitter and receiver have access to. It is assumed that the \ac{BS} and all users have access to perfect instantaneous \ac{CSI}. It is further assumed that users can perform perfect error-free \ac{SIC} to remove decoded signals from the received signal. In the following, we introduce the system model of the general mixed coding \ac{MU}-\ac{LP} system, displayed in \figref{fig:SDMA1HybridFigure}. Later we will derive the obfuscation-only and \ac{WT}-coding only results as special cases.

\subsection{Obfuscation \& Deobfuscation}
The system employs a protocol-specific reversible obfuscation function to transform the message $\mathbf{m}_k$ of each user $k$ into a ciphertext $\boldsymbol{\chi}_k$, utilizing a shared secret. Additionally, a protocol specific \ac{SI} $\boldsymbol{i}_k$ is produced, which must be transmitted securely. Examples for obfuscation  functions include conventional cryptographic \emph{computationally} secure \ac{AE} \cite[Ch.~9]{boneh2023}, \emph{information theoretically} secure \ac{OTP} encryption \cite{Gungor2012} or network coding \cite{Tajbakhsh2018}. Protocol examples of \ac{SI} could include a sequential key update \cite[Sec.~III]{Bloss2025}, random bits for an offline key reservoir \cite{Gungor2012}, or subsets of a network encoded message~\cite{Tajbakhsh2018}. 

After reception, the obfuscation  can be reversed by utilizing the received \ac{SI} $\hat{\boldsymbol{i}}_k$ and the received ciphertext $\hat{\boldsymbol{\chi}}_{k}$ to obtain $\hat{\mathbf{m}}_k$. Other ciphertexts $\hat{\boldsymbol{\chi}}_{k'}$ intended for users $k' \in \mathcal{K} \setminus \{k\}$ remain indecipherable to user $k$. This follows because the required \ac{SI} $\hat{\boldsymbol{i}}_{k'}$ is assumed to be protected by \ac{WT} coding and hence inaccessible to user $k$. In the system models in \figref{fig:SDMA1HybridFigure}, \figref{fig:SDMAOnlyObfs} and \figref{fig:SDMAOnlyPLS} all red operations enclosed by solid black borders are feasible to the eavesdropper, while dashed red borders are infeasible. \looseness=-1

\subsection{Transmit and Receive Procedure}
After preprocessing, $\boldsymbol{\chi}_k$ is split into the private and confidential parts $\mathbf{m}_{\mathrm{p},k}$ and $\mathbf{m}_{\upchi,\mathrm{\varsigma},k}$, respectively. Thereafter, $\mathbf{m}_{\upchi,\mathrm{\varsigma},k}$ is combined with the secret update $\boldsymbol{i}_k$ to form the confidential message $\mathbf{m}_{\mathrm{\varsigma},k}$. This message split is only made to arbitrary distribute $\boldsymbol{\chi}_k$ to either message, it is not required for security. 
The system then encodes the obfuscated private messages $\mathbf{m}_{\mathrm{p},k}$ without secrecy requirements using a regular encoder. The confidential messages $\mathbf{m}_{\mathrm{\varsigma},k}$ are encoded along a randomly generated private seed $\mathbf{r}_k$ using a stochastic \ac{WT} encoder. This produces the respective private streams $s_{\mathrm{p},k}$ and secure streams $s_{\mathrm{\varsigma},k}$. The stream attributes are summarized in \tabref{tab:streamSum}.
\begin{table}[t]
    \vspace*{2mm}
    \centering
    \caption{Mixed Coding - User $k$ Stream Comparison}
    \label{tab:streamSum}
    \setlength{\tabcolsep}{4pt}
    \renewcommand{\arraystretch}{1.2}
    \begin{tabular}{|l|c|c|c|c|}
        \hline
        \textbf{Stream} & \textbf{Name} & \textbf{Encoding} & \textbf{Secured by}   & \textbf{Carries}\\
        \hline
        \textbf{Private} & $s_{\mathrm{p},k}$
        & Regular
        & Obfuscation  & Obfuscated data \\
        \hline
        \textbf{Confidential} & $s_{\mathrm{\varsigma},k}$
        & \ac{WT}
        & \ac{WT} coding & Obfuscated data + SI \\
        \hline
    \end{tabular}
\end{table}

Note that $\mathbf{s} = [\mathbf{s}_{\mathrm{p}}, \mathbf{s}_{\mathrm{\varsigma}}]= [s_{\mathrm{p},1}, \, \ldots, \, s_{\mathrm{p},K},s_{\mathrm{\varsigma},1}, \, \ldots, \, s_{\mathrm{\varsigma},K}]^\mathrm{T}$ denotes the symbol vector for a single channel use, with source signals that are assumed to be uncorrelated and of unit power such that $\mathbb{E}[\mathbf{s}\mathbf{s}^H]= \mathbf{I}$. The \ac{BS} then maps $\mathbf{s}$ to the $N$ transmit antennas by means of an \ac{MU}–\ac{LP} precoder 
\begin{equation}
    \label{eq:xt_PLS}
    \! \mathbf{x} \!=\! \mathbf{W}_{\mathrm{p}} \mathbf{s}_{\mathrm{p}} \!+ \mathbf{W}_{\mathrm{\varsigma}} \mathbf{s}_{\mathrm{\varsigma}} \!= \! \sum_{k\in \mathcal{K}} \! \mathbf{w}_{\mathrm{p}, k} s_{\mathrm{p}, k} +\! \sum_{k\in \mathcal{K}} \! \mathbf{w}_{\mathrm{\varsigma}, k} \, s_{\mathrm{\varsigma}, k} \! \in \!\mathbb{C}^{N\times 1} \!\!
\end{equation}
where $[\mathbf{W}_{\mathrm{p}}, \mathbf{W}_{\mathrm{\varsigma}}] = [\mathbf{w}_{\mathrm{p},1}, \ldots, \mathbf{w}_{\mathrm{p},K},\mathbf{w}_{\mathrm{\varsigma},1}, \ldots, \mathbf{w}_{\mathrm{\varsigma},K}]$ denotes the precoder matrix, with  $\mathbf{w}_{\mathrm{p},k}$ and $\mathbf{w}_{\mathrm{\varsigma},k}$ denoting the unsecured private and confidential precoders serving user $k$, respectively. The signal is transmitted over a narrowband flat-fading \ac{AWGN} channel, with a channel coefficient $\mathbf{h}_k\in\mathbb{C}^{N\times 1}$. Each user $k$ then receives the signal $y_k = \mathbf{h}^H_k \mathbf{x} + n_k,$ witg $n_k \sim \mathcal{CN}(0, \sigma_{n,k}^2) $ describes the noise at user $k$, with $\sigma_{\mathrm{n},k}^2 = \sigma_{\mathrm{n}}^2, \,\forall k \, \in \, \mathcal{K}$. 

At reception, user $k$ first decodes the private message $\hat{\mathbf{m}}_{\mathrm{p},k}$ from $y_k$. Thereafter, user $k$ is assumed to re-encode and re-precode the signal. Then the channel $\mathbf{h}_k^H$ is re-applied to remove the contribution of $\hat{\mathbf{m}}_{\mathrm{p},k}$ from $y_k$ via \ac{SIC}. This yields the intermediate signal $y'_{k} = y_k - \mathbf{h}^H_k \mathbf{w}_{\mathrm{p},k} \hat{s}_{\mathrm{p},k}$ from which user $k$ decodes the confidential message $\hat{\mathbf{m}}_{\mathrm{\varsigma},k}$. Finally, the user message $\hat{\mathbf{m}}_{k}$ is reconstructed by reversing the splitting and combining steps of the transmitter, then deobfuscating  $\boldsymbol{\chi}_k$ with knowledge of $\hat{\boldsymbol{i}}_k$. 

\subsection{Achievable Rates}
To obtain tractable rate expressions, we adopt the Gaussian signaling model over an \ac{AWGN} channel. Under this assumption, we denote the maximum achievable rate at which user $k$ can reliably decode a stream $s_{\epsilon}$ as
\begin{equation} 
    R_{\mathrm{\epsilon},k} = \log_2\left(1 + \gamma_{\mathrm{\epsilon},k} \right) \quad \forall \; \mathrm{\epsilon} \, \in \{p,\varsigma\} \label{eq:ChannelCap}.
\end{equation}
with the \acp{SINR} 
\begin{subequations}
\begin{align}
    \gamma_{\mathrm{p},k} & \triangleq \frac{|\mathbf{h}_{k}^{H}\mathbf{w}_{\mathrm{p},k}|^2}{\sum_{j\in\mathcal{K} \setminus \{k\}} |\mathbf{h}_{k}^{H}\mathbf{w}_{\mathrm{p},j}|^2 + \sum_{i\in\mathcal{K}} |\mathbf{h}_{k}^{H}\mathbf{w}_{\mathrm{\varsigma},i}|^2 + \sigma_n^2},\\
    \gamma_{\mathrm{\varsigma},k} & \triangleq \frac{|\mathbf{h}_{k}^{H}\mathbf{w}_{\mathrm{\varsigma},k}|^2}{\sum_{j\in\mathcal{K} \setminus \{k\}} |\mathbf{h}_{k}^{H}\mathbf{w}_{\mathrm{p},j}|^2 + \sum_{i\in\mathcal{K} \setminus \{k\}} |\mathbf{h}_{k}^{H}\mathbf{w}_{\mathrm{\varsigma},i}|^2 + \sigma_n^2}.
\end{align}
\end{subequations}
for the private stream $s_{\mathrm{p},k}$ and confidential stream $s_{\mathrm{\varsigma},k}$. Additionally, an unintended user $k'\in \mathcal{K} \setminus \{k\}$ may eavesdrop on the confidential stream $s_{\mathrm{\varsigma},k}$ of user $k$ with
\begin{equation}
    \gamma_{\mathrm{\varsigma},k \leftarrow {k'}} \triangleq \frac{|\mathbf{h}_{k'}^{H}\mathbf{w}_{\mathrm{\varsigma},{k}}|^2}{\sum_{i \in\mathcal{K} \setminus \{k,k'\}} |\mathbf{h}_{k'}^{H}\mathbf{w}_{\mathrm{\varsigma},i}|^2 + \sigma_n^2}.
\end{equation}
where we conservatively assume that any user is able to obtain the signal $y^\ast$, by decoding and canceling \emph{all} private streams $s_{\mathrm{p},1}, \ldots, s_{\mathrm{p},K}$, as they are not explicitly secured by \ac{WT} coding. Under this architecture, the confidential stream $s_{\mathrm{\varsigma},k}$ may achieve the secrecy rate $R_{\mathrm{s},k} = \big[\, R_{\mathrm{\varsigma}, k} - R_{\mathrm{\varsigma}, k \leftarrow k^{\ast}} \,\big]^+\!$, where $[x]^{+} = \max\{x,0\}$ and $k^{\ast}$ denotes the most capable eavesdropper, who achieves the maximum eavesdropping rate ($\max_{ k' \in \set{K} \setminus \{k\}} \{R_{\mathrm{\varsigma}, k \leftarrow k'}\}$) \cite{Wyner1975}. Accordingly, the achievable secure data rate for user $k$ under is given by
\begin{equation}
    R^{\mathrm{mix}}_{\mathrm{d},k} =  R_{\mathrm{p},k} + R_{\mathrm{s},k} - \rho^{\mathrm{req}}_{\mathrm{i},k},
\end{equation}
where $\rho^{\mathrm{req}}_{\mathrm{i},k},$ denotes the \ac{SI} requirement of user $k$ and $\boldsymbol{\rho}^{\mathrm{req}}_{\mathrm{i}} = [\rho^{\mathrm{req}}_{\mathrm{i},1}, \, \ldots, \, \rho^{\mathrm{req}}_{\mathrm{i},K}]^\mathsf{T}$ the requirements of all users.

\subsection{Special case: only-obfuscation  and only-\ac{WT} coding}
\begin{figure}[!t]%
    \centering%
    \captionsetup{justification=centering,labelsep=space}%
    \includegraphics[width=\columnwidth]{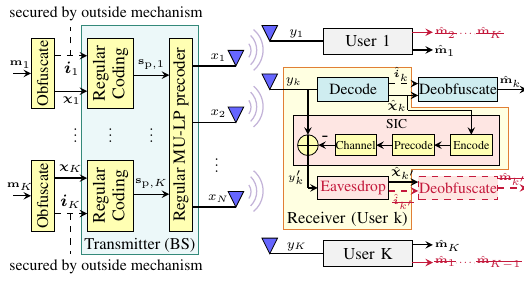}%
    \vspace{-1.5em}%
    \caption{System model for only-obfuscation  \ac{MU}-\ac{LP}}%
    \label{fig:SDMAOnlyObfs}%
\end{figure}

\begin{figure}[!t]%
    \centering%
    \captionsetup{justification=centering,labelsep=space}%
    \includegraphics[width=\columnwidth]{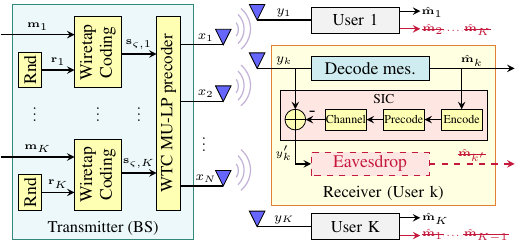}%
    \vspace{-0.5em}%
    \caption{System model for only-\ac{WT} coding \ac{MU}-\ac{LP}}%
    \label{fig:SDMAOnlyPLS}%
\end{figure}

Both special cases restrict the precoder selection and the expected \ac{SI} overhead. The obfuscation-only \ac{MU}-\ac{LP} system is visualized in \figref{fig:SDMAOnlyObfs}. In this system, only regularly encoded private streams are transmitted, such that $\mathbf{W}_{\mathrm{\varsigma}} = \mathbf{0}$. As a result, this system effectively corresponds to a regular \ac{MU}-\ac{LP} system without secrecy requirements, as described in \cite{Bjoernson2014}. But, this also means that $\mathbf{i}_k$ must be securely transferred to user $k$ using other means. The most common approach is to apply costly asymmetric cryptography to establish a shared key between participants, and then use efficient symmetric obfuscation  to protect the data \cite[Sec.~11.1, Ex.~11.9]{boneh2023}. As a result, we model the additional overhead required to establish $\mathbf{i}_k$, using an additional penalty term $\rho^{\mathrm{req}}_{\mathrm{a},k}$ to the data rate
\begin{equation}
    R^{\mathrm{obfs}}_{\mathrm{d},k} =  R_{\mathrm{p},k} - \rho^{\mathrm{req}}_{\mathrm{i},k} - \rho^{\mathrm{req}}_{\mathrm{a},k}, 
\end{equation}
where $\boldsymbol{\rho}^{\mathrm{req}}_{\mathrm{a}} = [\rho^{\mathrm{req}}_{\mathrm{a},1}, \, \ldots, \, \rho^{\mathrm{req}}_{\mathrm{a},K}]^\mathsf{T}$ denotes the additional protocol overhead required in the obfuscation-only scenario. In practical systems, this could correspond to overhead introduced by asymmetric key establishment. The \ac{WT}-only \ac{MU}-\ac{LP} system is visualized in \figref{fig:SDMAOnlyPLS}. If all data is secured via \ac{WT} coding, we obtain $\mathbf{W}_{\mathrm{p}} = \mathbf{0}$ and set $\boldsymbol{\rho}^{\mathrm{req}}_{\mathrm{i}} = \mathbf{0}$. The achievable rate is then $R^{\mathrm{wtc}}_{\mathrm{d},k} = R_{\mathrm{s},k}$.

\section{Optimization Problem}\label{sec:optproblem}
Based on this definition, we formulate the non-convex optimization problem to determine the optimal precoding matrices $\mathbf{W}{\mathrm{p}}$ and $\mathbf{W}{\mathrm{\varsigma}}$ for the proposed mixed-coding \ac{MU}-\ac{LP} system:
\begin{subequations} 
\label{eq:Prob_MixCode} \small
    \begin{align}
        & \mathcal{P}_{\mathrm{mix}} \!\!:  \max_{\mathbf{W}_{\mathrm{p}}, \mathbf{W}_{\mathrm{\varsigma}}} \sum_{k \in \mathcal{K}}  R^{\mathrm{mix}}_{\mathrm{d},k} 
        \label{eq:MixCode_objFunc} \\
        & \hspace{\PZEROsubjectToSpaceOptprob} \textrm{s.t.} \;\; R^{\mathrm{mix}}_{\mathrm{d},k}  \ge \rho^{\mathrm{req}}_{k}, \quad \forall \; k \in \mathcal{K},\,\label{eq:MixCode_QoS} \\
        & \hspace{\PZEROconstraintsSpaceOptprob} R_{\mathrm{s},k} \ge \rho^{\mathrm{req}}_{\mathrm{i},k}, \qquad \forall \; k \in \mathcal{K},\label{eq:MixCode_SecrecyQoS} \\
        & \hspace{\PZEROconstraintsSpaceOptprob} \|\mathbf{W}_{\mathrm{p}}\|_{\mathrm{F}}^2 + \|\mathbf{W}_{\mathrm{\varsigma}}\|_{\mathrm{F}}^2 \leq P_{\max}.\label{eq:MixCode_const_TXPwr}
    \end{align}
\end{subequations}
where $\rho^{\mathrm{req}}_{k}$ denotes the overall rate requirement of user $k$ and $\|\cdot\|_{\mathrm{F}}$ denotes the Frobenius norm. Accordingly, constraint \eqref{eq:MixCode_QoS} ensures the overall \ac{QoS} of each user. Constraint \eqref{eq:MixCode_SecrecyQoS} ensures that the achieved secrecy rate $R_{\mathrm{s},k}$ is sufficient to transport the \ac{SI} in the protected confidential stream. Constraint \eqref{eq:MixCode_const_TXPwr} limits the maximum transmit power to $P_{\max}$. We collect all \ac{QoS} requirements into the vector $\boldsymbol{\rho}^{\mathrm{req}} = [\rho^{\mathrm{req}}_{1}, \, \ldots, \, \rho^{\mathrm{req}}_{k}]^\mathsf{T}$. Additionally, we define the precoder optimization problem for the obfuscation-only \ac{MU}-{LP} architecture, as
\begin{subequations} 
\label{eq:Prob_OnlyObfs} \small
    \begin{align}
        \mathcal{P}_{\mathrm{obfs}} & \!\!:  \max_{\mathbf{W}_{\mathrm{p}}} \sum_{k \in \set{K}}  R^{\mathrm{obfs}}_{\mathrm{d},k} 
        \label{eq:SDMACrypt_objFunc} \\
        & \hspace{\PZEROsubjectToSpaceOptprob} \textrm{s.t.} \;\; R^{\mathrm{obfs}}_{\mathrm{d},k} \ge \rho^{\mathrm{req}}_{k} \qquad \forall \; k \in \mathcal{K},\label{eq:SDMACrypt_QoS} \\
        & \hspace{\PZEROconstraintsSpaceOptprob} \|\mathbf{W}_{\mathrm{p}}\|_{\mathrm{F}}^2 \leq P_{\max} \label{eq:SDMACrypt_const_TXPwr}.
    \end{align}
\end{subequations} 
where constraint \eqref{eq:SDMACrypt_QoS} ensures the overall \ac{QoS} of each user and constraint \eqref{eq:SDMACrypt_const_TXPwr} limits the maximum transmit power to $P_{\max}$. Finally, we define the optimization problem for the conventional \ac{MU}-{LP} \ac{WT} coding-only system, as 
\begin{subequations} 
\label{eq:Prob_OnlyPLS} \small
    \begin{align}
        \mathcal{P}_{\mathrm{wtc}} & \!\!:  \max_{\mathbf{W}_{\mathrm{\varsigma}}} \sum_{k \in \set{K}} R^{\mathrm{wtc}}_{\mathrm{d},k}
        \label{eq:SDMAPLS_objFunc} \\
        & \hspace{\PZEROsubjectToSpaceOptprob} \textrm{s.t.} \;\; R^{\mathrm{wtc}}_{\mathrm{d},k} \ge \rho^{\mathrm{req}}_{k} \qquad \forall \; k \in \mathcal{K},\label{eq:SDMAPLS_QoS} \\
        & \hspace{\PZEROconstraintsSpaceOptprob}  \|\mathbf{W}_{\mathrm{\varsigma}}\|_{\mathrm{F}}^2 \leq P_{\max} \label{eq:SDMAPLS_const_TXPwr}, 
    \end{align}
\end{subequations} 
where constraint \eqref{eq:SDMAPLS_QoS} ensures the overall \ac{QoS} of each user and constraint \eqref{eq:SDMAPLS_const_TXPwr} limits the transmit power to $P_{\max}$. 

All three problems are non-convex and cannot be directly solved using CVX. To reformulate the problem, we employ the same \ac{SCA}-based method presented in our previous work \cite{Bloss2025}. This method first reformulates the problems using auxiliary variables to isolate non-convex expressions. The isolated non-convex expressions are then linearized using the first-order Taylor approximation, while preserving feasibility. Due to this linearization, the method converges only to a stationary point, not the global optimum. The underlying method is well established in the precoder-optimization literature (see, e.g., Refs.~63, 124, 127, and 128 in \cite{Clerckx2023}). For this reason, the detailed presentation is omitted at this point but a complete derivation for problem \eqref{eq:Prob_MixCode} is provided in the Appendix. 
\section{Simulation Results}\label{sec:ResultsBeampattern}
\begin{figure}[t]
  \centering
    \includegraphics{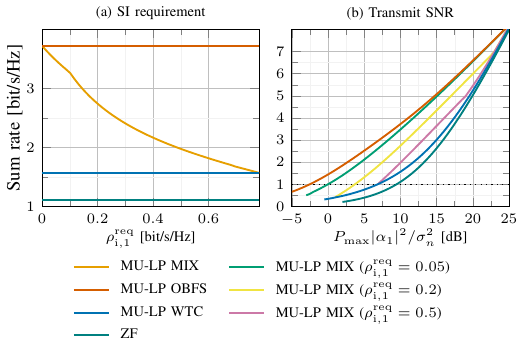}
    \vspace{-1.5em}
    \caption{Effects of SI requirement and SNR on sum-rate ($\rho^{\mathrm{req}}_{\mathrm{i},1} = \rho^{\mathrm{req}}_{\mathrm{i},2}$).} 
    \label{fig:bridging}
\end{figure}
In the following simulations, we compare the three specified methods and \ac{WT}-coded closed-form \ac{ZF} \cite{Bjoernson2014}, such that $R^{\mathrm{zf}}_{\mathrm{d},k} = R_{\mathrm{s},k}$. We assume a two user scenario ($K=2$) with two transmit antennas ($N=2$) corresponding to the overloaded regime \cite{Clerckx2023}, where mixed coding is expected to offer the greatest benefit. The reason is that with an increasing number of antennas per user, the precoder can more effectively separate users spatially, thereby reducing the rate reduction associated with wiretap coding, as shown in \cite{Bloss2025}. For $N \gg K$, closed-form \ac{ZF} can even become close to optimal \cite{Bjoernson2014}. 

We further omit the penalty terms from the objective functions, since these capture communication overhead introduced by specific obfuscation functions and would require further assessment for realistic parameterization. Moreover, these penalties are not central to investigating the influence of the minimum secrecy requirement $\boldsymbol{\rho}^{\mathrm{req}}_{\mathrm{i},k}$ in constraint \eqref{eq:MixCode_SecrecyQoS} and the individual beampatterns of the coding layers. We therefore define the objective functions as
\begin{equation}
    R^{\mathrm{mix}}_{\mathrm{d},k} =  R_{\mathrm{p},k} + R_{\mathrm{s},k}, \; R^{\mathrm{wtc}}_{\mathrm{d},k} = R_{\mathrm{s},k}, \; R^{\mathrm{obfs}}_{\mathrm{d},k} =  R_{\mathrm{p},k}.
\end{equation}
The channel is defined as a narrowband, far-field geometric \ac{LoS} channel based on the \ac{UPW} model defined in \cite[Sec.~II-A]{Lu2024}. We denote the channel of user $k$ as 
\begin{equation}
\mathbf{h}_k = \alpha_k\mathbf{a}(\theta_k),
\qquad
\alpha_k = \frac{\lambda}{4\pi r_k}
e^{j\frac{2\pi}{\lambda}r_k},
\end{equation}
where $\boldsymbol{\theta} = [\theta_1,\ldots,\theta_K]^\mathrm{T},\boldsymbol{\theta}\in[-90^\circ,90^\circ]^K$ denotes the angle of departures relative to the array broadside, and $\mathbf{r} = [r_1, \ldots, r_K]^\mathrm{T}, $ the propagation distance of user $k$, relative to the array center at $(0,0)$. The transmit \ac{SNR} is defined relative to the first user as $\eta = P_{\max}|\alpha_1|^2 / \sigma_n^2$.The transmit array response is characterized by $\mathbf{a}(\theta_k)$, while $\alpha_k \in \mathbb{C}$ incorporates the free-space propagation attenuation and phase. Assuming a \ac{ULA}, we denote $\mathbf{a}(\theta_k)$ as
\begin{equation} 
    \mathbf{a}(\theta) =
    \left[
    e^{j\frac{\pi d}{\lambda}(N-1)\sin\theta},
    \,\ldots,\,
    e^{-j\frac{\pi d}{\lambda}(N-1)\sin\theta}
    \right]^T. 
\end{equation}
where $d$ describes the inter-element spacing. 

The optimization problems reveal that the mixed coding in Problem \eqref{eq:Prob_MixCode}, can freely allocate transmit power between $\mathbf{W}_{\mathrm{p}}$ and $\mathbf{W}_{\mathrm{\varsigma}}$ as long as constraint \eqref{eq:MixCode_SecrecyQoS} is fulfilled. Thus, higher values of $\boldsymbol{\rho}^{\mathrm{req}}_{\mathrm{i}}$, impose tighter constraints on the precoder, which must ensure that more data is be delivered securely by \ac{WT} coding.  A low secrecy requirement $\boldsymbol{\rho}^{\mathrm{req}}_{\mathrm{i}}$ results in an increased sum-rate, while a higher secrecy requirement $\boldsymbol{\rho}^{\mathrm{req}}_{\mathrm{i}}$ reduces the overall sum-rate. The border cases also resemble the two special cases $\boldsymbol{\rho}^{\mathrm{req}}_{\mathrm{i}}=\mathbf{0}$ corresponds to regular coding, described in Problem \eqref{eq:Prob_OnlyObfs}, while the maximum achievable $\boldsymbol{\rho}^{\mathrm{req}}_{\mathrm{i}}$ corresponds to full \ac{WT} coding, described in Problem \eqref{eq:Prob_OnlyPLS}. 

\figref{fig:bridging}(a) illustrates this connection, by considering a scenario with symmetric requirements: $\mathbf{r} = [\qty{100}{\metre}, \qty{100}{\metre}]^{\mathrm{T}}$, $\boldsymbol{\theta} = [0^{\circ}, 8^{\circ}]^{\mathrm{T}}$, $\boldsymbol{\rho}^{\mathrm{req}} = [0.1, 0.1]^{\mathrm{T}}$, $\rho^{\mathrm{req}}_{\mathrm{i},1} = \rho^{\mathrm{req}}_{\mathrm{i},2}$ and $\eta = \qty{10}{\decibel}$ and $\sigma^2_n = 1$. In this scenario, we observe that the maximum serviceable $\boldsymbol{\rho}^{\mathrm{req}}_{\mathrm{i}}$ occurs at $\boldsymbol{\rho}^{\mathrm{req}}_{\mathrm{i}} \approx [0.784 , 0.784]$, when $\sum R^{\mathrm{mix}}_{\mathrm{d},k} = \sum R^{\mathrm{wtc}}_{\mathrm{d},k}$. Additionally, \figref{fig:bridging}(b) substantiates this claim. A low secrecy requirement of $\rho^{\mathrm{req}}_{\mathrm{i},1} = 0.05$ leads to higher sum-rates, and converges to the sum-rate of regular coding for high \ac{SNR}. Also, high secrecy requirement of $\rho^{\mathrm{req}}_{\mathrm{i},1} = 0.5$ lead to problem infeasibility at low \ac{SNR} values. Performance differences between the coding strategies are also more pronounced at low \ac{SNR}. In summary, the simulation results indicate that mixed coding can provide a performance benefit depending on the secrecy requirement $\boldsymbol{\rho}^{\mathrm{req}}_{\mathrm{i}}$. 

To understand where this improvement stems from, we analyze the beampatterns and rate allocation of the individual streams. Following \cite[Eq.~(54)]{Lu2024}, we define the beampattern of stream $\epsilon$ intended for user $k$ as
\begin{equation}
    {B}_{\mathrm{\epsilon},k}(\theta) = 10\log_{10}\left(\frac{|\mathbf{a}^{H}(\theta)\mathbf{w}_{\mathrm{\epsilon},k}|^2}{N P_{\mathrm{max}}}
    \right),\quad \mathrm{\epsilon} \in \{\mathrm{p},\varsigma\}.
\end{equation}
Rather than normalizing each stream individually, we normalize by the theoretical maximum directional power $N P_{\mathrm{max}}$ of a single stream. This preserves the power allocation between beams, and ensures ${B}_{\mathrm{\epsilon},k} \! \leq \! \qty{0}{\decibel}$. Additionally, we choose $d  = \lambda/2$ and $f_c  =\qty{8}{\giga\hertz}$. The requirements are set to $\boldsymbol{\rho}^{\mathrm{req}}_{\mathrm{i}} \! = [0.05, 0.05]^{\mathrm{T}}$, $\boldsymbol{\rho}^{\mathrm{req}} \! = [0.1, 0.1]^{\mathrm{T}}$ and the \ac{SNR} to $\qty{10}{\decibel}$.

\begin{figure*}[!t]%
    \centering%
    \captionsetup{justification=centering,labelsep=space}%
    \includegraphics[width=\textwidth]{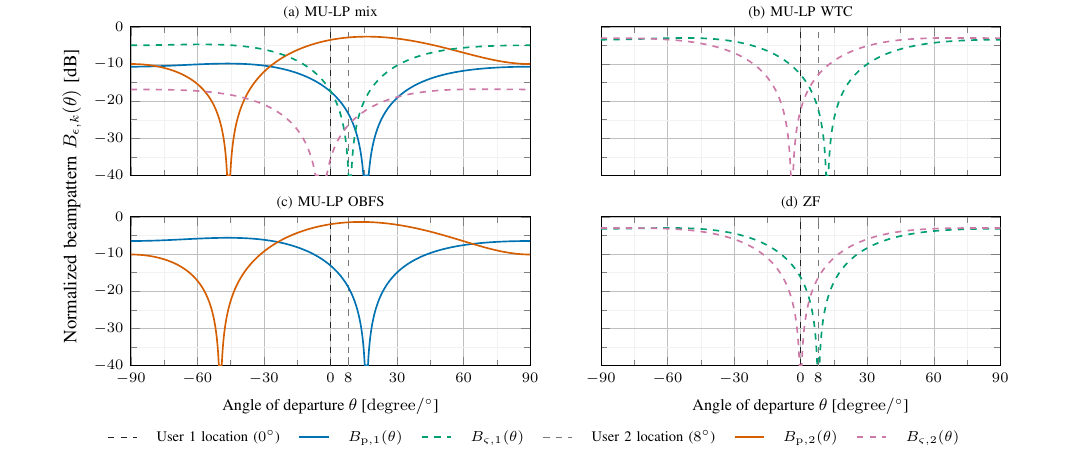}%
    \vspace{-0.5em}%
    \caption{Beampatterns of individual streams with $\theta_1=0^\circ$ and
    $\theta_2=8^\circ$, normalized by $N P_{\mathrm{max}}$.}
    \label{fig:beampatternComp}%
    \vspace{-1em}%
\end{figure*}

The beampatterns are shown in \figref{fig:beampatternComp}, while the numerical rates are shown in \tabref{tab:beampatternRates}. Examining \tabref{tab:beampatternRates} reveals two distinct rate allocation strategies. Both \ac{MU}-\ac{LP} MIX and \ac{MU}-\ac{LP} OBFS allocate most rate towards a single user, while the other user is served only at their minimum \ac{QoS}. Additionally, \ac{MU}-\ac{LP} MIX only allocates the minimum secrecy rate $R_{\mathrm{s},1},  R_{\mathrm{s},2} = 0.05$, to satisfy $\boldsymbol{\rho}^{\mathrm{req}}_{\mathrm{i}}$. On the other hand, \ac{MU}-\ac{LP} WTC and \ac{ZF} allocate equal rates to both users. 

Relating these rate allocations to the beampatterns reveals the underlying precoding strategies. \ac{MU}-\ac{LP} WTC minimizes inter-user interference, since the secrecy-rate objective in \eqref{eq:SDMAPLS_objFunc} penalizes information leakage. This results in a beampattern similar to that of \ac{ZF}, which aims to eliminate inter-user interference entirely. \tabref{tab:beampatternRates} also reveals that complete interference cancellation is not optimal in the considered scenario, as $R^{\mathrm{wtc}}_{\mathrm{d},1}>R^{\mathrm{zf}}_{\mathrm{d},1}$ and $R^{\mathrm{wtc}}_{\mathrm{d},2}>R^{\mathrm{zf}}_{\mathrm{d},2}$. In contrast, \ac{MU}-\ac{LP} OBFS adopts an \ac{MRT}-like strategy \cite{Bjoernson2014}, maximizing beamforming gain towards user 2 at the expense of increased interference.
Mixed coding combines both strategies: $\mathbf{W}{\mathrm{\varsigma}}$ follows the minimize-leakage strategy of \ac{ZF}, while $\mathbf{W}{\mathrm{p}}$ follows an \ac{MRT}-like approach to maximize beamforming gain at user 2, accepting increased interference at user 1. 

Together, the observations from \figref{fig:bridging} and \figref{fig:beampatternComp} highlight the main synergy between mixed coding and precoding in the \ac{MU}-\ac{LP} architecture: Mixed coding isolates the secrecy requirement of \ac{WT} coding into a separate stream, enabling the precoder to shape the beams of each coding layer individually. Furthermore, the results in \figref{fig:beampatternComp} indicate that $|\mathbf{W}{\mathrm{p}}|{\mathrm{F}}^2 > |\mathbf{W}{\mathrm{\varsigma}}|{\mathrm{F}}^2$, while the secrecy rate remains at its minimum requirement, suggesting that the precoder limits the resources allocated to the more restrictive \ac{WT}-coded stream.

\begin{table}[t]
    \vspace*{2mm}
    \centering
    \caption{Beampattern - Achievable Rates [bit/s/Hz]}
    \label{tab:beampatternRates}
    \setlength{\tabcolsep}{5pt}
    \renewcommand{\arraystretch}{1.2}
    \begin{tabular}{|l|c|c|c|c|c|c|}
        \hline
        \textbf{Method} & $R_{\mathrm{d},1}$ & $R_{\mathrm{d},2}$ & $R_{\mathrm{p},1}$
        & $R_{\mathrm{s},1}$ & $R_{\mathrm{p},2}$ & $R_{\mathrm{s},2}$ \\
        \hline
        \textbf{MU-LP MIX}  & 0.100 & 3.376 & 0.050 & 0.050 & 3.326 & 0.050 \\
        \hline
        \textbf{MU-LP OBFS} & 0.100 & 3.621 & 0.100 & --- & 3.621 & ---\\
        \hline
        \textbf{MU-LP WTC}  & 0.784 & 0.784 & --- & 0.784 & --- & 0.784 \\
        \hline
        \textbf{ZF}         & 0.556 & 0.556 & --- & 0.556 & --- & 0.556 \\
        \hline
    \end{tabular}
\end{table}

\begin{figure*}[t]
  \centering
  \includegraphics[width=\textwidth]{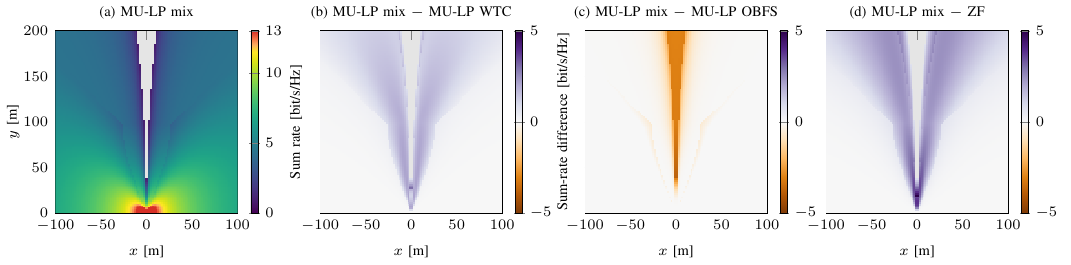}
  \vspace{-1.5em}
  \caption{MU-LP MIX sum-rate heatmap (a) and differences against
    \ac{WT}-only coding (b), obfuscation-based regular coding (c),
    and ZF (d).}
  \label{fig:heatmapComp}
  \vspace{-2em}
\end{figure*}

Finally, we utilize the geometric channel model to investigate how the performance gains of mixed coding depend on the user positioning. The position of user~1 is assumed to be fixed at $\theta_1 = 0^{\circ}$ and $r_1 = \qty{100}{\metre}$, while the position of user~2 is varied over a $100 \times 100$ grid spanning $x \in [-\qty{100}{\metre}, \qty{100}{\metre}]$ and $y \in [\qty{0}{\metre}, \qty{200}{\metre}]$, with a spacing of $\qty{2}{\metre}$ between sample points. At each position, we evaluate the achievable sum rate of \ac{MU}-\ac{LP} with mixed coding, resulting in the heatmap shown in \figref{fig:heatmapComp}(a). Gray sample points indicate scenarios in which the \ac{SCA} algorithm determined the problem to be infeasible. 

Subtracting the corresponding sum-rate heatmaps of the other coding strategies results in the differential heatmaps in \figref{fig:heatmapComp}(b)-(d). Particularly relevant is \figref{fig:heatmapComp}(b), which reveals the spatial configurations in which mixed coding outperforms \ac{WT} coding. Gray sample points indicate scenarios in which the \ac{SCA} algorithm determined both problem to be infeasible. If only one problem was deemed infeasible, we assumed a sum-rate of $0$ for this method. The diagrams show that that mixed coding achieves the largest sum-rate gains over \ac{WT} coding when both users are located at similar angles. In our geometric far-field model, such configurations result in highly aligned channels, making spatial separation costly in terms of achievable sum rate. Mixed coding alleviates this problem by requiring only the \ac{SI} to be protected by \ac{WT} coding.

\vspace{-0.5em}
\section{Conclusion}\label{sec:conclusion}
This work presented a \ac{MU}-\ac{LP}-based system architecture that utilizes an abstract obfuscation function to combine regular and wiretap coding. We formulated the corresponding precoder optimization problems for mixed coding and the two individual coding strategies, with the mixed-coding formulation abstracting the obfuscation function through a minimum secrecy rate constraint. Numerical simulations using a geometric two-user \ac{LoS} channel revealed that mixed coding can bridge the gap between regular and wiretap coding, with the minimum secrecy rate requirement determining the operating point between the two strategies. The subsequent beampattern analysis revealed an important synergy between mixed coding and precoding: Mixed coding isolates the secrecy requirement to a separate stream, allowing the precoder to shape the beams of the two coding layers individually. The spatial analysis further revealed that mixed coding achieves its largest sum-rate gains over exclusive wiretap coding when users are located at similar angles relative to the transmitter.
However, the simulations were limited to a two-user scenario and did not account for the communication overhead introduced by the obfuscation function. Future work could therefore extend the analysis to scenarios with more than two users and incorporate realistic estimates of the communication overhead to assess its impact on the achievable performance. \looseness=-1
\vspace{-0.5em}
\section*{Acknowledgment} \small
\footnotesize{Large parts of the simulation and visualization code were generated using GPT-6 Astra from OpenAI Codex (v0.160.1) on reasoning settings \emph{extra high}. The model was given access to the content of \secref{sec:SystemModel} and \secref{sec:optproblem}, as well as the appendix of this work and simulation code from our previous work \cite{Bloss2025}. The code was verified by the authors and Claude code independently.}

\footnotesize{The authors gratefully acknowledge the computing time made available to them on the high-performance computer Romeo at the NHR Center of TU Dresden. This center is jointly supported by the Federal Ministry of Research, Technology and Space of Germany and the state governments participating in the NHR (www.nhr-verein.de/unsere-partner).}

\acresetall
\appendix
\onecolumn
\definecolor{redCol}{HTML}{dd1c77}\normalsize
\subsection{Introduction: Successive Convex Approximation} \label{app:secSCAINtro}
As discussed in \secref{sec:optproblem}, the precoder optimization problems \eqref{eq:Prob_MixCode}, \eqref{eq:Prob_OnlyObfs} and \eqref{eq:Prob_OnlyPLS} are all non-convex and cannot directly be solved by CVX. For this reason, we reformulate the problem using \ac{SCA} \cite{Razaviyayn2014}. \Ac{SCA} has been adopted in the wireless communications literature \cite{Xia2022, Xia2024Journal, Bloss2025, Tervo2017, Zhou2022, Mao2018SCA, Cui2023, Ibrahim2020, AlObiedollah2019} to find solutions to non-convex problems. The main idea is to construct and successively solve tractable convex subproblems that approximate the original non-convex problem tightly around an expansion point. After each iteration, the solution of the current convex subproblem is used as the expansion point for the next iteration. The algorithm terminates when the objective value improves only marginally between consecutive iterations.

In the following \secref{app:secCeonvexSubProb}, we start from the original non-convex problem \eqref{eq:Prob_MixCode} and derive the convexified subproblem. To achieve this, we introduce auxiliary variables to isolate non-convex expressions and obtain a difference-of-convex structure, with only a few terms remaining problematic and causing the constraints to be non-convex. These terms are then linearized using the first-order Taylor approximation. Finally, \secref{app:secIterAlgo} presents the algorithm for successively solving the convex subproblems, followed by a brief convergence argument based on \cite{Bloss2025} and the description of the chosen initial expansion points.

\subsection{Derivation of Convexified Subproblem for $\mathcal{P}_{\mathrm{mix}}$} \label{app:secCeonvexSubProb}
For this derivation, we assume that each user requires a nonzero secrecy rate $\rho^{\mathrm{req}}_{\mathrm{i},k} > 0 \; \forall \, k \, \in \, \mathcal{K}$ and nonnegative \ac{QoS} $\rho^{\mathrm{req}}_{k} \geq 0$ requirements.
Throughout the derivation, we highlight the terms in \textcolor{redCol}{magenta} that cause non-convexity in the constraint. The starting point is Problem \eqref{eq:Prob_MixCode}:
\begin{subequations} 
\label{eq:App_Problem_SDMA_1SIC} \small
    \begin{align} \small
        & \mathcal{P}_{\mathrm{mix}} \!\!:  \max_{\mathbf{W}_{\mathrm{p}}, \mathbf{W}_{\mathrm{\varsigma}}} \sum_{k \in \set{K}} \left(\mathcolor{redCol}{R_{\mathrm{p},k}+R_{\mathrm{s},k}}-\rho^{\mathrm{req}}_{\mathrm{i},k}\right) \label{eqapp:v2_initial_objective}\\
        & \hspace{\PZEROsubjectToSpaceOptprob} \textrm{s.t.} \;\; \mathcolor{redCol}{R_{\mathrm{p},k}+R_{\mathrm{s},k}}-\rho^{\mathrm{req}}_{\mathrm{i},k} \ge \rho^{\mathrm{req}}_{k}, \qquad \forall \; k \in \set{K}, \label{eqapp:v2_initial_QoS}\\
        & \hspace{\PZEROconstraintsSpaceOptprob} \mathcolor{redCol}{R_{\mathrm{s},k}} \ge \rho^{\mathrm{req}}_{\mathrm{i},k}, \qquad \forall \; k \in \set{K},\label{eqapp:v2_initial_key_rate}\\
        & \hspace{\PZEROconstraintsSpaceOptprob} \sum_{k \in \mathcal{K}} ||\mathbf{w}_{\mathrm{p}, k}||^2 + \sum_{k \in \mathcal{K}} ||\mathbf{w}_{\mathrm{\varsigma}, k}||^2   \leq P_{\max}. \label{eqapp:v2_power_constraint} 
    \end{align}
\end{subequations}
Initially, the \ac{QoS} constraint \eqref{eqapp:v2_initial_QoS} and the minimum secrecy rate constraint \eqref{eqapp:v2_initial_key_rate} are non-convex. Additionally, the rate terms in the objective \eqref{eqapp:v2_initial_objective} are non-concave, which is required for maximization problems. The transmit-power constraint \eqref{eqapp:v2_power_constraint} is convex, and does not require further treatment. First we can make use of the strictly positive secrecy-rate requirement ($\rho^{\mathrm{req}}_{\mathrm{i},k} > 0 \; \forall \, k \, \in \, \mathcal{K}$). Recall that the secrecy rate of user $k$ is defined as
\begin{equation}
R_{\mathrm{s},k} = \left[ R_{\mathrm{\varsigma},k} - \max_{k'\in\mathcal K\setminus\{k\}} R_{\mathrm{\varsigma},k\leftarrow k'} \right]^+,
\end{equation}
where $[x]^{+} = \max\{x,0\}$ and $k^{\ast}$ denotes the most capable eavesdropper, who achieves the maximum eavesdropping rate ($\max_{ k' \in \set{K} \setminus \{k\}} \{R_{\mathrm{\varsigma}, k \leftarrow k'}\}$). At any feasible point, constraint \eqref{eqapp:v2_initial_key_rate} requires $R_{\mathrm{s},k}\geq\rho^{\mathrm{req}}_{\mathrm{i},k}>0$. The rate difference inside the positive-part operation must therefore be strictly positive, and thus the $[\cdot]^{+}$ operation can be omitted. Further, we observe that the constraint
\begin{equation}
R_{\mathrm{\varsigma},k} - \max_{k'\in\mathcal K\setminus\{k\}} R_{\mathrm{\varsigma},k\leftarrow k'} \geq \rho^{\mathrm{req}}_{\mathrm{i},k} \quad  \forall \; k \in \set{K}
\end{equation}
can be equivalently expressed as
\begin{equation}
R_{\mathrm{\varsigma},k} - R_{\mathrm{\varsigma},k\leftarrow k'} \geq \rho^{\mathrm{req}}_{\mathrm{i},k}, \quad \forall \; k \in \set{K}, \; k' \in \set{K} \setminus \{k\},
\end{equation}
which makes use of the fact that when the minimum secrecy-rate requirement holds for any potential eavesdropper $\forall k'\in\mathcal K\setminus\{k\}$, then it automatically holds for the most capable eavesdropper $k^{\ast}$.

In the next step, the rate expressions are isolated. We therefore introduce the variables $\boldsymbol{\rho} = \{\rho_{\mathrm{d},k},\,
\rho_{\mathrm{s},k},\,\rho_{\mathrm{p},k}\in\mathbb R_{\geq0} \mid k\in\mathcal K\}$:
\begin{subequations} 
    \begin{align}
        & \max_{\mathbf{W}_{\mathrm{p}}, \mathbf{W}_{\mathrm{\varsigma}}, \boldsymbol{\rho}} \sum_{k \in \set{K}} \rho_{\mathrm{d},k} \\
        & \hspace{\PZEROsubjectToSpaceOptprob} \textrm{s.t.} \;\; \rho_{\mathrm{d},k} \ge \rho^{\mathrm{req}}_{k}, \qquad \forall \; k \in \set{K}, \label{eqapp:v2_rho_dk_const_convex}\\
        & \hspace{\PZEROconstraintsSpaceOptprob}  \rho_{\mathrm{s},k}  \ge \rho^{\mathrm{req}}_{\mathrm{i},k}, \qquad \forall \; k \in \set{K},\label{eqapp:v2_rho_sk_const_convex}\\
        & \hspace{\PZEROconstraintsSpaceOptprob}  \rho_{\mathrm{d},k} \ge 0, \qquad \forall \; k \in \set{K},\label{eqapp:v2_rho_dk_two_const_convex}\\
        & \hspace{\PZEROconstraintsSpaceOptprob} \sum_{k \in \mathcal{K}} ||\mathbf{w}_{\mathrm{p}, k}||^2 + \sum_{k \in \mathcal{K}} ||\mathbf{w}_{\mathrm{\varsigma}, k}||^2   \leq P_{\max},\\
        & \hspace{\PZEROconstraintsSpaceOptprob}  \rho_{\mathrm{s},k}  +  \rho_{\mathrm{p},k} - \rho^{\mathrm{req}}_{\mathrm{i},k}  \geq \rho_{\mathrm{d},k}, \qquad \forall \; k \in \set{K},\label{eqapp:v2_rho_dk_rate_convex}\\     
         & \hspace{\PZEROconstraintsSpaceOptprob} \mathcolor{redCol}{\log_2( 1+\gamma_{\mathrm{p},k})} \geq \rho_{\mathrm{p},k} \qquad \forall k\in\mathcal{K},  \label{eqapp:v2_priv_reli_nonconvex}\\
         & \hspace{\PZEROconstraintsSpaceOptprob} \mathcolor{redCol}{\log_2( 1+\gamma_{\mathrm{\varsigma},k} ) - \log_2 ( 1+\gamma_{\mathrm{\varsigma},k \leftarrow {k'}} )} \geq  \rho_{\mathrm{s},k} \qquad \forall \; \subalign{&k \in \set{K},\\&k' \in \set{K} \setminus \{k\}}.\label{eqapp:v2_sec_reli_nonconvex}
    \end{align}
\end{subequations}
The objective is now linear and constraints \eqref{eqapp:v2_rho_dk_const_convex}, \eqref{eqapp:v2_rho_sk_const_convex}, \eqref{eqapp:v2_rho_dk_two_const_convex}, \eqref{eqapp:v2_rho_dk_rate_convex} are affine. Only the private-rate constraint \eqref{eqapp:v2_priv_reli_nonconvex} and the secrecy-rate constraint \eqref{eqapp:v2_sec_reli_nonconvex} remain non-convex. The reason is that the \ac{SINR} expressions $\gamma_{\mathrm{p},k}$, $\gamma_{\mathrm{\varsigma},k}$ and $\gamma_{\mathrm{\varsigma},k \leftarrow {k'}}$ are quadratic-over-quadratic expressions in $\mathbf{W}_{\mathrm{p}}$ and $\mathbf{W}_{\mathrm{\varsigma}}$ and thus the $\log_2 ( 1+\gamma)$ expression is generally non-concave.

To further isolate the non-convex part, we next isolate the \ac{SINR} expressions, by introducing the variables:
$\boldsymbol{\alpha}=\{\alpha_{\mathrm{p},k},\,
\alpha_{\mathrm{\varsigma},k},\,
\alpha_{\mathrm{\varsigma},k\leftarrow k'}\in\mathbb R_{\geq0}
\mid k\in\mathcal K,\,k'\in\mathcal K\setminus\{k\}\}$:
\allowdisplaybreaks[2] 
\begin{subequations} 
    \begin{align}
        & \max_{\mathbf{W}_{\mathrm{p}}, \mathbf{W}_{\mathrm{\varsigma}}, \boldsymbol{\rho}, \boldsymbol{\alpha}} \sum_{k \in \set{K}} \rho_{\mathrm{d},k} \\
        & \hspace{\PZEROsubjectToSpaceOptprob} \textrm{s.t.} \;\; \rho_{\mathrm{d},k} \ge \rho^{\mathrm{req}}_{k}, \qquad \forall \; k \in \set{K}, \\
        & \hspace{\PZEROconstraintsSpaceOptprob}  \rho_{\mathrm{s},k}  \ge \rho^{\mathrm{req}}_{\mathrm{i},k}, \qquad \forall \; k \in \set{K},\\
        & \hspace{\PZEROconstraintsSpaceOptprob}  \rho_{\mathrm{d},k} \ge 0, \qquad \forall \; k \in \set{K},\\
        & \hspace{\PZEROconstraintsSpaceOptprob} \sum_{k \in \mathcal{K}} ||\mathbf{w}_{\mathrm{p}, k}||^2 + \sum_{k \in \mathcal{K}} ||\mathbf{w}_{\mathrm{\varsigma}, k}||^2   \leq P_{\max},\\
        & \hspace{\PZEROconstraintsSpaceOptprob}  \rho_{\mathrm{s},k}  +  \rho_{\mathrm{p},k} - \rho^{\mathrm{req}}_{\mathrm{i},k}  \geq \rho_{\mathrm{d},k}, \qquad \forall \; k \in \set{K}, \\
         & \hspace{\PZEROconstraintsSpaceOptprob} \log_2( 1+\alpha_{\mathrm{p},k}) \geq \rho_{\mathrm{p},k} \qquad \forall k\in\mathcal{K}, \label{eqapp:v2_alpha_private_rate}\\
         & \hspace{\PZEROconstraintsSpaceOptprob} \log_2( 1+\alpha_{\mathrm{\varsigma},k} ) \mathcolor{redCol}{- \log_2 ( 1+\alpha_{\mathrm{\varsigma},k \leftarrow {k'}} )} \geq  \rho_{\mathrm{s},k}  \qquad \forall \; \subalign{&k \in \set{K},\\&k' \in \set{K} \setminus \{k\}} \label{eqapp:v2_alpha_secrecy_rate}\\
         & \hspace{\PZEROconstraintsSpaceOptprob} \mathcolor{redCol}{\frac{|\mathbf{h}_{k}^{H}\mathbf{w}_{\mathrm{p},k}|^2}{\sum_{j\in\mathcal{K} \setminus \{k\}} |\mathbf{h}_{k}^{H}\mathbf{w}_{\mathrm{p},j}|^2 + \sum_{i\in\mathcal{K}} |\mathbf{h}_{k}^{H}\mathbf{w}_{\mathrm{\varsigma},i}|^2 + \sigma_n^2}} \geq \alpha_{\mathrm{p},k} \qquad \forall k\in\mathcal{K}, \label{eqapp:v2_priv_SINR_nonconvex} \\
         & \hspace{\PZEROconstraintsSpaceOptprob} \mathcolor{redCol}{\frac{|\mathbf{h}_{k}^{H}\mathbf{w}_{\mathrm{\varsigma},k}|^2}{\sum_{j\in\mathcal{K} \setminus \{k\}} |\mathbf{h}_{k}^{H}\mathbf{w}_{\mathrm{p},j}|^2 + \sum_{i\in\mathcal{K} \setminus \{k\}} |\mathbf{h}_{k}^{H}\mathbf{w}_{\mathrm{\varsigma},i}|^2 + \sigma_n^2}} \geq \alpha_{\mathrm{\varsigma},k} \qquad \forall k\in\mathcal{K}, \label{eqapp:v2_sigma_SINR_nonconvex}\\
         & \hspace{\PZEROconstraintsSpaceOptprob} \mathcolor{redCol}{\frac{|\mathbf{h}_{k'}^{H}\mathbf{w}_{\mathrm{\varsigma},{k}}|^2}{\sum_{i \in\mathcal{K} \setminus \{k,k'\}} |\mathbf{h}_{k'}^{H}\mathbf{w}_{\mathrm{\varsigma},i}|^2 + \sigma_n^2}} \leq \alpha_{\mathrm{\varsigma},k\leftarrow k'} \qquad \forall \; \subalign{&k \in \set{K},\\&k' \in \set{K} \setminus \{k\}} \label{eqapp:v2_eve_SINR_nonconvex}
    \end{align}
\end{subequations}
The private-rate constraint \eqref{eqapp:v2_alpha_private_rate} is now convex, since $1+ \alpha_{\mathrm p,k}$ is affine and thus $\log_2(1+\alpha_{\mathrm p,k})$ is concave in $\alpha_{\mathrm p,k}$. The secrecy-rate constraint \eqref{eqapp:v2_alpha_secrecy_rate} remains non-convex because the difference of logarithms is not generally concave. The quadratic-over-quadratic \ac{SINR} expressions are generally neither concave nor convex in $\mathbf{W}_{\mathrm{p}}$ and $\mathbf{W}_{\mathrm{\varsigma}}$. Consequently, constraints \eqref{eqapp:v2_priv_SINR_nonconvex}, \eqref{eqapp:v2_sigma_SINR_nonconvex}, and \eqref{eqapp:v2_eve_SINR_nonconvex} are non-convex.

In the next step, we further isolate the denominators of the \ac{SINR} expressions. We therefore introduce the variable set $\boldsymbol{\beta}=\{\beta_{\mathrm{p},k},\,
\beta_{\mathrm{\varsigma},k},\, \beta_{\mathrm{\varsigma},k\leftarrow k'}\in\mathbb R_{>0} \mid k\in\mathcal K,\,k'\in\mathcal K\setminus\{k\}\}$:
\begin{subequations} 
\label{eq:App_Problem_final_reform} \small
    \begin{align}
        & \max_{\mathbf{W}_{\mathrm{p}}, \mathbf{W}_{\mathrm{\varsigma}}, \boldsymbol{\rho}, \boldsymbol{\alpha}, \boldsymbol{\beta}} \sum_{k \in \set{K}} \rho_{\mathrm{d},k} \\
        & \hspace{\PZEROsubjectToSpaceOptprob} \textrm{s.t.} \;\; \rho_{\mathrm{d},k} \ge \rho^{\mathrm{req}}_{k}, \qquad \forall \; k \in \set{K}, \\
        & \hspace{\PZEROconstraintsSpaceOptprob}  \rho_{\mathrm{s},k}  \ge \rho^{\mathrm{req}}_{\mathrm{i},k}, \qquad \forall \; k \in \set{K},\\
        & \hspace{\PZEROconstraintsSpaceOptprob}  \rho_{\mathrm{d},k} \ge 0, \qquad \forall \; k \in \set{K},\\
        & \hspace{\PZEROconstraintsSpaceOptprob} \sum_{k \in \mathcal{K}} ||\mathbf{w}_{\mathrm{p}, k}||^2 + \sum_{k \in \mathcal{K}} ||\mathbf{w}_{\mathrm{\varsigma}, k}||^2   \leq P_{\max},\\
        & \hspace{\PZEROconstraintsSpaceOptprob}  \rho_{\mathrm{s},k}  +  \rho_{\mathrm{p},k} - \rho^{\mathrm{req}}_{\mathrm{i},k}  \geq \rho_{\mathrm{d},k}, \qquad \forall \; k \in \set{K}, \\ 
        & \hspace{\PZEROconstraintsSpaceOptprob} \log_2( 1+\alpha_{\mathrm{p},k})  \geq \rho_{\mathrm{p},k} \quad \forall k\in\mathcal{K}, \label{eqapp:v2_private_rate_alpha_convex}\\
        & \hspace{\PZEROconstraintsSpaceOptprob} \log_2( 1+\alpha_{\mathrm{\varsigma},k} ) \mathcolor{redCol}{-\log_2 ( 1+\alpha_{\mathrm{\varsigma},k \leftarrow {k'}} )} \geq  \rho_{\mathrm{s},k}   \quad \forall \; \subalign{&k \in \set{K},\\&k' \in \set{K} \setminus \{k\}} \label{eqapp:v2_highlighted_alpha_sigmak}\\
        & \hspace{\PZEROconstraintsSpaceOptprob} \mathcolor{redCol}{\frac{|\mathbf{h}_{k}^{H}\mathbf{w}_{\mathrm{p},k}|^2}{\beta_{\mathrm{p},k}}} \geq \alpha_{\mathrm{p},k} \qquad \forall k\in\mathcal{K}, \label{eqapp:v2_priv_SINR_split_nonconvex} \\
        & \hspace{\PZEROconstraintsSpaceOptprob} \sum_{j\in\mathcal{K} \setminus \{k\}} |\mathbf{h}_{k}^{H}\mathbf{w}_{\mathrm{p},j}|^2 + \sum_{i\in\mathcal{K}} |\mathbf{h}_{k}^{H}\mathbf{w}_{\mathrm{\varsigma},i}|^2 + \sigma_n^2 \leq \beta_{\mathrm{p},k} \qquad \forall k\in\mathcal{K}, \label{eqapp:v2_priv_SINR_split_convex}\\         
        & \hspace{\PZEROconstraintsSpaceOptprob} \mathcolor{redCol}{\frac{|\mathbf{h}_{k}^{H}\mathbf{w}_{\mathrm{\varsigma},k}|^2}{\beta_{\mathrm{\varsigma},k}}} \geq \alpha_{\mathrm{\varsigma},k} \qquad \forall k\in\mathcal{K}, \label{eqapp:v2_sigma_SINR_split_nonconvex}\\
        & \hspace{\PZEROconstraintsSpaceOptprob} \sum_{j\in\mathcal{K} \setminus \{k\}} |\mathbf{h}_{k}^{H}\mathbf{w}_{\mathrm{p},j}|^2 + \sum_{i\in\mathcal{K} \setminus \{k\}} |\mathbf{h}_{k}^{H}\mathbf{w}_{\mathrm{\varsigma},i}|^2 + \sigma_n^2 \leq \beta_{\mathrm{\varsigma},k} \qquad \forall k\in\mathcal{K}, \label{eqapp:v2_sigma_SINR_split_convex}\\ 
        & \hspace{\PZEROconstraintsSpaceOptprob} \frac{|\mathbf{h}_{k'}^{H}\mathbf{w}_{\mathrm{\varsigma},{k}}|^2}{\beta_{\mathrm{\varsigma},k\leftarrow k'}} \leq \alpha_{\mathrm{\varsigma},k\leftarrow k'} \qquad \forall \; \subalign{&k \in \set{K},\\&k' \in \set{K} \setminus \{k\}}, \label{eqapp:v2_eve_SINR_split_convex_frac}\\
        & \hspace{\PZEROconstraintsSpaceOptprob} \mathcolor{redCol}{\sum_{i \in\mathcal{K} \setminus \{k,k'\}} |\mathbf{h}_{k'}^{H}\mathbf{w}_{\mathrm{\varsigma},i}|^2} + \sigma_n^2 \geq \beta_{\mathrm{\varsigma},k\leftarrow k'} \qquad \forall \; \subalign{&k \in \set{K},\\&k' \in \set{K} \setminus \{k\}}, \label{eqapp:v2_eve_SINR_split_nonconvex_sum} 
    \end{align}
\end{subequations}
Constraints \eqref{eqapp:v2_priv_SINR_split_convex} and \eqref{eqapp:v2_sigma_SINR_split_convex} contain the denominators of the previous private and confidential \ac{SINR} expressions. These constraints are convex because they contain sums of convex quadratic terms and a constant, bounded above by affine variables. Constraint \eqref{eqapp:v2_eve_SINR_split_nonconvex_sum}, on the other hand, is generally non-convex because the convex quadratic sum is bounded below by an affine variable. The fractional parts of the \ac{SINR} expressions reverse this pattern. Constraints \eqref{eqapp:v2_priv_SINR_split_nonconvex} and \eqref{eqapp:v2_sigma_SINR_split_nonconvex} contain the fractional parts of the previous private and confidential \ac{SINR} constraints and are generally non-convex. This is because they contain convex quadratic-over-linear functions bounded below by affine variables. The inequality direction is reversed for the eavesdropper constraint: the convex quadratic-over-linear function is bounded above by an affine variable, and thus constraint \eqref{eqapp:v2_eve_SINR_split_convex_frac} is convex.

Finally, the reformulation is finished and the constraints \eqref{eqapp:v2_highlighted_alpha_sigmak}, \eqref{eqapp:v2_priv_SINR_split_nonconvex}, \eqref{eqapp:v2_sigma_SINR_split_nonconvex}, \eqref{eqapp:v2_eve_SINR_split_nonconvex_sum} remain generally non-convex. The terms responsible for this non-convexity are still highlighted in \textcolor{redCol}{magenta}. To obtain a convex subproblem that can be solved using CVX, we next linearize the problematic terms.

The linearization is performed similarly to the \ac{CCP} construction described by Lipp and Boyd~\cite{Lipp2016}, by retaining the constant and first-order terms of the complex Taylor expansion. However, the procedure is adjusted to encompass real-valued functions $f(\mathbf{x}_1,\ldots,\mathbf{x}_K)$ with respect to the complex vector variables $\mathbf{x}_1,\ldots,\mathbf{x}_K$ \cite[Eq.~(5.40)]{Hjorungnes2011}.  Around the expansion points $(\mathbf{x}_1^{[n]},\ldots,\mathbf{x}_K^{[n]})$, this gives
\begin{equation}
\begin{aligned}
f(\mathbf{x}_1,\ldots,\mathbf{x}_K) \approx f\bigl(\mathbf{x}_1^{[n]},\ldots,\mathbf{x}_K^{[n]}\bigr) + & \sum_{k=1}^K \left[ \left( \nabla_{\mathbf{x}_k}  f\bigl(\mathbf{x}_1^{[n]},\ldots,\mathbf{x}_K^{[n]}\bigr) \right)^T \Delta \mathbf{x}_k  + \left( \nabla_{\mathbf{x}_k^*} f\bigl(\mathbf{x}_1^{[n]},\ldots,\mathbf{x}_K^{[n]}\bigr) \right)^T \Delta \mathbf{x}_k^*\right],
\end{aligned}
\end{equation}
with $\Delta \mathbf{x}_k \triangleq \mathbf{x}_k-\mathbf{x}_k^{[n]}$ and $\Delta \mathbf{x}^{\ast}_k \triangleq \mathbf{x}^{\ast}_k-(\mathbf{x}^{[n]}_k)^{\ast}$. Here, $\nabla_{\mathbf x_k}f$ and $\nabla_{\mathbf x_k^{\ast}}f$ denote column vectors that collect the Wirtinger derivatives with respect to the $\mathbf x_k$ and $\mathbf x_k^{\ast}$ components, respectively \cite[Definition~2.2]{Hjorungnes2011}. If $f(\mathbf{x}_1,\ldots,\mathbf{x}_K)$ is convex, then its first-order Taylor approximation provides a lower bound \cite[Sec.~1.3]{Lipp2016}, such that 
\begin{equation}
f(\mathbf{x}_1,\ldots,\mathbf{x}_K) \geq f\bigl(\mathbf{x}_1^{[n]},\ldots,\mathbf{x}_K^{[n]}\bigr) + \sum_{k=1}^K \left[ \left( \nabla_{\mathbf{x}_k}  f\bigl(\mathbf{x}_1^{[n]},\ldots,\mathbf{x}_K^{[n]}\bigr) \right)^T  \Delta \mathbf{x}_k  +  \left( \nabla_{\mathbf{x}_k^{\ast}} f\bigl(\mathbf{x}_1^{[n]},\ldots,\mathbf{x}_K^{[n]}\bigr) \right)^T \Delta \mathbf{x}_k^{\ast} \right].
\end{equation}
If $f(\mathbf{x}_1,\ldots,\mathbf{x}_K)$ is concave, then an upper bound 
\begin{equation}
f(\mathbf{x}_1,\ldots,\mathbf{x}_K) \leq f\bigl(\mathbf{x}_1^{[n]},\ldots,\mathbf{x}_K^{[n]}\bigr) + \sum_{k=1}^K \left[ \left( \nabla_{\mathbf{x}_k}  f\bigl(\mathbf{x}_1^{[n]},\ldots,\mathbf{x}_K^{[n]}\bigr) \right)^T  \Delta \mathbf{x}_k  +  \left( \nabla_{\mathbf{x}_k^{\ast}} f\bigl(\mathbf{x}_1^{[n]},\ldots,\mathbf{x}_K^{[n]}\bigr) \right)^T \Delta \mathbf{x}_k^{\ast} \right]
\end{equation}
is provided instead. To calculate the first-order Taylor approximation, we need the first-order derivatives of the problematic terms in constraints \eqref{eqapp:v2_highlighted_alpha_sigmak}, \eqref{eqapp:v2_priv_SINR_split_nonconvex}, \eqref{eqapp:v2_sigma_SINR_split_nonconvex} and \eqref{eqapp:v2_eve_SINR_split_nonconvex_sum}. The derivatives are given by 
\begin{subequations}
    \begin{align}
    \frac{d}{d\alpha}\log_2(1+\alpha) & = \frac{\log_2(e)}{(1+\alpha)}= \frac{1}{(1+\alpha)\ln(2)},\label{eqapp:v2_log_derivative}\\
    \frac{\partial}{\partial\mathbf{w}}\frac{|\mathbf{h}^H \mathbf{w}|^2}{\beta} & = \frac{(\mathbf{w}^H \mathbf{h} \mathbf{h}^H)^T}{\beta},\label{eqapp:v2_quad_over_lin_derivative_w} \\
    \frac{\partial}{\partial\mathbf{w}^{\ast}}\frac{|\mathbf{h}^H \mathbf{w}|^2}{\beta} & = \frac{\mathbf{h} \mathbf{h}^H \mathbf{w}}{\beta}, \label{eqapp:v2_quad_over_lin_derivative_wstar}\\
    \frac{\partial}{\partial\beta}\frac{|\mathbf{h}^H \mathbf{w}|^2}{\beta} & = - \frac{|\mathbf{h}^H \mathbf{w}|^2}{\beta^2},\label{eqapp:v2_quad_over_lin_derivative_beta}\\
    \frac{\partial}{\partial \mathbf{w}_{i}} |\mathbf{h}^{H}\mathbf{w}_{i}|^{2} &= (\mathbf{w}_{i}^{H}\mathbf{h}\mathbf{h}^{H})^T, \label{eqapp:v2_quad_abs_derivative_w}\\
    \frac{\partial}{\partial \mathbf{w}_{i}^{\ast}} |\mathbf{h}^{H}\mathbf{w}_{i}|^{2} &= \mathbf{h}\mathbf{h}^{H}\mathbf{w}_{i}.\label{eqapp:v2_quad_abs_derivative_wstar}
    \end{align}
\end{subequations}
Using \eqref{eqapp:v2_log_derivative}, we approximate constraint \eqref{eqapp:v2_highlighted_alpha_sigmak} as
\begin{equation}
    \begin{split}
        &\log_2(1+\alpha_{\mathrm{\varsigma},k}) - \log_2\!\left(1+\alpha_{\mathrm{\varsigma},k\leftarrow k'}^{[n]}\right) - \frac{\alpha_{\mathrm{\varsigma},k\leftarrow k'} -\alpha_{\mathrm{\varsigma},k\leftarrow k'}^{[n]}}{\left(1+\alpha_{\mathrm{\varsigma},k\leftarrow k'}^{[n]}\right)\ln 2} \geq \rho_{\mathrm{s},k}, \quad \forall k\in\mathcal K, \; k' \in\set{K} \setminus\{k\}.
        \label{eqapp:v2_solved_secrate_sigma_stream}
    \end{split}
\end{equation}
Following the previous argument, we recognize that the original function $-\log_2(1+\alpha_{\mathrm{\varsigma},k\leftarrow k'})$ is convex. As a result, the first order Taylor approximation provides an affine lower bound, meaning that 
\begin{equation}
-\log_2\!\left(1+\alpha_{\mathrm{\varsigma},k\leftarrow k'}^{[n]}\right) -\frac{\alpha_{\mathrm{\varsigma},k\leftarrow k'} -\alpha_{\mathrm{\varsigma},k\leftarrow k'}^{[n]}}{\left(1+\alpha_{\mathrm{\varsigma},k\leftarrow k'}^{[n]}\right)\ln2} \leq -\log_2(1+\alpha_{\mathrm{\varsigma},k\leftarrow k'}).
\end{equation}
Consequently, the approximated secrecy-rate expression is a lower bound on the original expression. Any point satisfying the approximated constraint \eqref{eqapp:v2_solved_secrate_sigma_stream} also satisfies the original secrecy-rate constraint \eqref{eqapp:v2_highlighted_alpha_sigmak}.
The non-convex constraints \eqref{eqapp:v2_priv_SINR_split_nonconvex} and \eqref{eqapp:v2_sigma_SINR_split_nonconvex} are linearized using the derivatives \eqref{eqapp:v2_quad_over_lin_derivative_w}, \eqref{eqapp:v2_quad_over_lin_derivative_wstar} and \eqref{eqapp:v2_quad_over_lin_derivative_beta}. Both constraints are based on the quadratic-over-linear function 
\begin{equation}
f(\mathbf{w},\beta)=\frac{|\mathbf{h}^H\mathbf{w}|^2}{\beta}, \qquad \beta>0,
\end{equation}
which is jointly convex in $(\mathbf{w},\beta)$. We utilize \eqref{eqapp:v2_quad_over_lin_derivative_w}, \eqref{eqapp:v2_quad_over_lin_derivative_wstar}
and \eqref{eqapp:v2_quad_over_lin_derivative_beta} to derive the linearization, leading to the affine lower bound
\begin{equation}
    \label{eqapp:v2_solved_frac_terms_general} 
    \begin{aligned}
        \frac{|\mathbf{h}^H\mathbf{w}|^2}{\beta} 
        & \geq \frac{|\mathbf{h}^H\mathbf{w}^{[n]}|^2}{\beta^{[n]}} + \frac{2\Re\!\left\{(\mathbf{w}^{[n]})^H\mathbf{h}\mathbf{h}^H(\mathbf{w}-\mathbf{w}^{[n]})\right\}}{\beta^{[n]}} - \frac{|\mathbf{h}^H\mathbf{w}^{[n]}|^2}{(\beta^{[n]})^2}(\beta-\beta^{[n]}) \\
        &= \frac{2\Re\!\left\{(\mathbf{w}^{[n]})^H\mathbf{h}\mathbf{h}^H\mathbf{w}\right\}}{\beta^{[n]}} - \frac{|\mathbf{h}^H\mathbf{w}^{[n]}|^2\beta}{(\beta^{[n]})^2}.
    \end{aligned}
\end{equation}
Applying the first-order approximation in equation \eqref{eqapp:v2_solved_frac_terms_general} to the fractional terms in constraints \eqref{eqapp:v2_priv_SINR_split_nonconvex} and \eqref{eqapp:v2_sigma_SINR_split_nonconvex} yields
\begin{subequations} 
    \begin{align}
    &\frac{2\Re\!\left\{(\mathbf{w}_{\mathrm{p},k}^{[n]})^H \mathbf{h}_k\mathbf{h}_k^H \mathbf{w}_{\mathrm{p},k}\right\}}{\beta_{\mathrm{p},k}^{[n]}}  - \frac{|\mathbf{h}_k^H\mathbf{w}_{\mathrm{p},k}^{[n]}|^2 \cdot \beta_{\mathrm{p},k}}{(\beta_{\mathrm{p},k}^{[n]})^2} \geq \alpha_{\mathrm{p},k}, \quad \forall k \in \set{K}, \label{eqapp:v2_solved_private_SINR} \\
    &\frac{2\Re\!\left\{(\mathbf{w}_{\mathrm{\varsigma},k}^{[n]})^H \mathbf{h}_k\mathbf{h}_k^H \mathbf{w}_{\mathrm{\varsigma},k}\right\}}{\beta_{\mathrm{\varsigma},k}^{[n]}}  - \frac{|\mathbf{h}_k^H\mathbf{w}_{\mathrm{\varsigma},k}^{[n]}|^2 \cdot \beta_{\mathrm{\varsigma},k}}{(\beta_{\mathrm{\varsigma},k}^{[n]})^2} \geq \alpha_{\mathrm{\varsigma},k}, \quad \forall k \in \set{K}. \label{eqapp:v2_solved_sigma_SINR}    
    \end{align}
\end{subequations}
Finally, we utilize \eqref{eqapp:v2_quad_abs_derivative_w} and \eqref{eqapp:v2_quad_abs_derivative_wstar} to linearize constraint \eqref{eqapp:v2_eve_SINR_split_nonconvex_sum}, leading to
\begin{equation}
    \sum_{i \in\mathcal{K} \setminus \{k,k'\}} \left(2\Re\!\left\{(\mathbf{w}_{\mathrm{\varsigma},i}^{[n]})^H \mathbf{h}_{k'}\mathbf{h}_{k'}^H \mathbf{w}_{\mathrm{\varsigma},i}\right\} - |\mathbf{h}_{k'}^H\mathbf{w}_{\mathrm{\varsigma},i}^{[n]}|^2\right) + \sigma_n^2 \geq \beta_{\mathrm{\varsigma},k\leftarrow k'}, \quad \forall k\in \set{K},\; k'\in \set{K} \setminus\{k\}. \label{eqapp:v2_solved_eve_SINR_sum}
\end{equation}
The problematic sum of interference terms in constraint \eqref{eqapp:v2_eve_SINR_split_nonconvex_sum} is convex in $\mathbf{W}_{\mathrm{\varsigma}}$. The first-order Taylor approximation therefore provides an affine lower bound, such that
\begin{equation} 
\sum_{i\in\mathcal K\setminus\{k,k'\}}|\mathbf h_{k'}^H\mathbf w_{\mathrm{\varsigma},i}|^2+\sigma_n^2 \geq \sum_{i\in\mathcal K\setminus\{k,k'\}}\left(2\Re\!\left\{(\mathbf w_{\mathrm{\varsigma},i}^{[n]})^H\mathbf h_{k'}\mathbf h_{k'}^H\mathbf w_{\mathrm{\varsigma},i}\right\}-|\mathbf h_{k'}^H\mathbf w_{\mathrm{\varsigma},i}^{[n]}|^2\right)+\sigma_n^2. 
\end{equation}
Consequently, any point satisfying the approximated constraint \eqref{eqapp:v2_solved_eve_SINR_sum} also satisfies the original constraint \eqref{eqapp:v2_eve_SINR_split_nonconvex_sum}.

After linearizing the remaining non-convex terms, we obtain the convexified subproblem
\begin{subequations}
    \label{eqapp:v2_final_problem}
    \begin{align}
         \mathcal{P}^{\mathrm{final}}_{\mathrm{mix}}&\!:
        \max_{\mathbf{W}_{\mathrm{p}}, \mathbf{W}_{\mathrm{\varsigma}},\boldsymbol\rho,\boldsymbol\alpha,\boldsymbol\beta}
        \sum_{k\in\mathcal K}\rho_{\mathrm d,k}\\
        &\textrm{s.t.}\quad
        \eqref{eqapp:v2_power_constraint}, \eqref{eqapp:v2_rho_dk_const_convex}, \eqref{eqapp:v2_rho_sk_const_convex},
        \eqref{eqapp:v2_rho_dk_two_const_convex},
        \eqref{eqapp:v2_rho_dk_rate_convex},
        \eqref{eqapp:v2_private_rate_alpha_convex},
        \eqref{eqapp:v2_priv_SINR_split_convex}, \eqref{eqapp:v2_sigma_SINR_split_convex},
        \eqref{eqapp:v2_eve_SINR_split_convex_frac}, \eqref{eqapp:v2_solved_secrate_sigma_stream},
        \eqref{eqapp:v2_solved_private_SINR},
        \eqref{eqapp:v2_solved_sigma_SINR},
        \eqref{eqapp:v2_solved_eve_SINR_sum}.
    \end{align}
\end{subequations}
The subproblem has a linear objective and convex constraints for fixed expansion points, collected in the tuple 
\begin{equation} 
\label{eqapp:expansion_point_def}
\boldsymbol{\nu}^{[n]} = \left( \mathbf{W}_{\mathrm{p}}^{[n]}, \mathbf{W}_{\mathrm{\varsigma}}^{[n]}, \{\beta_{\mathrm{p},k}^{[n]}, \beta_{\mathrm{\varsigma},k}^{[n]}\}_{k\in\mathcal K}, \{\alpha_{\mathrm{\varsigma},k\leftarrow k'}^{[n]}\}_{k\in\mathcal K,\;k'\in\mathcal K\setminus\{k\}} \right). 
\end{equation}
Thus, we approximate the original problem $\mathcal{P}_{\mathrm{mix}}$ in \eqref{eq:App_Problem_SDMA_1SIC} by the convexified subproblem \eqref{eqapp:v2_final_problem}.
In the next part, we present the iterative algorithm used to repeatedly evaluate problem~\eqref{eqapp:v2_final_problem}, followed by a brief discussion about the convergence guarantees and the initial expansion points $\boldsymbol{\nu}^{[0]}$ used in this work.

\subsection{Algorithm, Convergence and Initial Expansion Point}  \label{app:secIterAlgo}
The proposed algorithm follows the successive convex approximation framework \cite[Algorithm~1]{Razaviyayn2014} and iteratively solves the corresponding convexified subproblem~\eqref{eqapp:v2_final_problem}. The proposed algorithm is presented in Algorithm~\ref{alg:successive_algo_mix} and terminates when the absolute difference between the objective values of two consecutive iterations falls below the threshold $\tau$, or when the maximum number of iterations $n_{\mathrm{max}}$ is reached. We use the solution of each convex subproblem directly as the next expansion point, corresponding to a step size of $\gamma=1$ in \cite[Algorithm~1]{Razaviyayn2014}. This step size was also used in our previous work \cite{Bloss2025} and by Xia et al.~\cite{Xia2022, Xia2024Journal}. The iteration index is denoted by $n$, while the expansion point at iteration $n$ is denoted by $\bm{\nu}^{[n]}$. Consequently, $\bm{\nu}^{[0]}$ denotes the initial expansion point.  

The performance requirements are defined as described in \secref{sec:SystemModel} and \secref{sec:optproblem}:
\begin{itemize}
    \item $\boldsymbol{\rho}^{\mathrm{req}} = [\rho^{\mathrm{req}}_{1}, \, \ldots, \, \rho^{\mathrm{req}}_{K}]^\mathsf{T}$: \Ac{QoS} / Minimum data rate requirements
    \item $\boldsymbol{\rho}^{\mathrm{req}}_{\mathrm{i}} = [\rho^{\mathrm{req}}_{\mathrm{i},1}, \, \ldots, \, \rho^{\mathrm{req}}_{\mathrm{i},K}]^\mathsf{T}$: Secrecy \Ac{QoS} / Minimum secrecy rate requirements (only relevant for mixed-coding)
\end{itemize}
For notational convenience, we define the objective value of iteration $n$ as $\mathrm{SR}^{[n]}$, such that
\begin{equation}
    \mathrm{SR}^{[n]}=\sum_{k\in\mathcal K}\rho_{\mathrm d,k}^{[n]}.
\end{equation}
Finally, we denote the optimal objective value of the current convex subproblem by $\mathrm{SR}^{\ast}$. From the solution attaining this value, we collect the auxiliary variables listed in \eqref{eqapp:expansion_point_def}, as well as the precoder matrices $\mathbf{W}_{\mathrm p}^{\ast}$ and $\mathbf{W}_{\mathrm{\varsigma}}^{\ast}$ into $\bm{\nu}^{\ast}$, which becomes the next expansion point.

\textit{Convergence}: We follow the same line of argument presented in \cite{Razaviyayn2014, Xia2022, Xia2024Journal, Bloss2025}. Provided that the first convex subproblem is feasible, the algorithm produces a non-decreasing sequence of objective values. As established in \secref{app:secCeonvexSubProb}, the linearizations preserve feasibility in the original problem \eqref{eq:App_Problem_SDMA_1SIC}. Since the first-order Taylor expansion is used, the approximated terms also equal the original terms at the expansion points. Therefore, the complete solution obtained in one iteration remains feasible in the convex subproblem of the following iteration. Thus, solving consecutive subproblems cannot decrease the objective value. Additionally, the achievable sum-rate is upper bounded by the transmit-power budget and the strictly positive noise. Consequently, the sequence of objective values converges \cite{Xia2022}. Note that convergence cannot be guaranteed if the initial subproblem is infeasible.

\textit{Initialization}: The solution obtained using \ac{SCA} can depend heavily on the initial expansion point \cite{Lipp2016}. We therefore run Algorithm~\ref{alg:successive_algo_mix} from multiple initial expansion points $\bm{\nu}^{[0]}$ and retain the result with the highest objective value. Note, that none of our initial expansion points are guaranteed to produce a feasible first convex subproblem. For all of these expansion points, we only select the initial precoders $\mathbf{W}^{[0]}_{\mathrm{p}}$ and $\mathbf{W}^{[0]}_{\mathrm{\varsigma}}$. The auxiliary variables $\alpha^{[0]}_{\mathrm{\varsigma},k\leftarrow k'}$, $\beta^{[0]}_{\mathrm{p},k}$ and $\beta^{[0]}_{\mathrm{\varsigma},k}$ are then initialized by extracting $\mathbf{w}_{\mathrm{p},k}$ from $\mathbf{W}^{[0]}_{\mathrm{p}}$ and $\mathbf{w}_{\mathrm{\varsigma},k}$ from $\mathbf{W}^{[0]}_{\mathrm{\varsigma}}$ and replacing the inequalities in constraints \eqref{eqapp:v2_eve_SINR_nonconvex}, \eqref{eqapp:v2_priv_SINR_split_convex} and \eqref{eqapp:v2_sigma_SINR_split_convex} with equalities. To initialize $\mathbf{W}^{[0]}_{\mathrm{p}}$ and $\mathbf{W}^{[0]}_{\mathrm{\varsigma}}$, we consider several different parameters: 1) Precoder structure, 2) Power allocation between private and confidential streams, 3) Power allocation between users, 4) Initial power scaling. 

\textit{Precoder Structure}: As shown by Björnson et al.~\cite{Bjoernson2014}, the optimal transmit precoder for regular coded \ac{MU}-\ac{LP} balances \ac{MRT} and \ac{ZF} precoding. Accordingly, we consider both structures for initializing the private streams. The confidential streams, on the other hand, are initialized either using \ac{ZF} directly or by using the precoders obtained by solving the \ac{WT}-coded \ac{MU}-\ac{LP} Problem \eqref{eq:Prob_OnlyPLS}. The \ac{WT}-coded \ac{MU}-\ac{LP} method itself is initialized using \ac{ZF} for the confidential streams.

\textit{Power allocation between private and confidential streams}: For mixed coding, we control how much of the power budget $P_{\max}$ is allocated to the private streams. Therefore, we introduce the variable $\delta_1\in[0,1)$, such that $\delta_1=0$ results in $\mathbf{W}^{[0]}_{\mathrm{p}} = \mathbf{0}$. Note that $\mathbf{W}^{[0]}_{\mathrm{p}} = \mathbf{0}$, effectively disables the private streams. For this reason, $\delta_1 = 1$ is not feasible as no power would be allocated to the confidential streams and since we assume strictly positive minimum secrecy-rate requirements, this would lead to an infeasible initial expansion point.

\textit{Power allocation between users}: Next, we want to pre-configure the power allocated to the precoders of each user. For the considered two-user scenario, we allocate $\delta_2$ to user~1 and $(1-\delta_2)$ to user~2. Together, with the power allocation between streams $\delta_1$, we obtain
\begin{equation} 
\|\mathbf{w}^{[0]}_{\mathrm{p},1}\|^2=\delta_2\delta_1P_{\max}, \qquad \|\mathbf{w}^{[0]}_{\mathrm{p},2}\|^2=(1-\delta_2)\delta_1P_{\max}, 
\end{equation}
and
\begin{equation}
\|\mathbf{w}^{[0]}_{\mathrm{\varsigma},1}\|^2=\delta_2(1-\delta_1)P_{\max}, \qquad \|\mathbf{w}^{[0]}_{\mathrm{\varsigma},2}\|^2=(1-\delta_2)(1-\delta_1)P_{\max}. 
\end{equation}
To obtain these powers while preserving the beam directions, we normalize each precoding vector to unit norm and multiply them by the square roots of their allocated powers. 

\textit{Initial power scaling}: Preliminary tests showed that it can be beneficial to initialize the precoders with a total transmit power below $P_{\max}$. We therefore introduce a power-scaling factor $\delta_3\in(0,1]$ and multiply both initial precoding matrices by $\sqrt{\delta_3}$. This scaling is applied after distributing the power to users and streams via $\delta_1$ and $\delta_2$. The resulting total initial transmit power is $\delta_3P_{\max}$.

Finally, we summarize all simulation parameters in \tabref{tab:simulation_Parameters}.

\begin{algorithm}
\caption{Iterative SCA-based algorithm for mixed-coding MU-LP precoder optimization}
\label{alg:successive_algo_mix}
\KwIn{$\tau$, \, $n_{\mathrm{max}}$, \, $\boldsymbol{\rho}^{\mathrm{req}}$, \, $\boldsymbol{\rho}^{\mathrm{req}}_{\mathrm{i}}$, \, $P_{\max}$, \, $\bm{\nu}^{[0]}$} 
$n \gets 0$\Comment*[r]{Initialize $n$}%
$\mathrm{SR}^{\ast} \gets -\infty$\Comment*[r]{Initialize optimal objective value}%
\Repeat{
$|\mathrm{SR}^{\ast} - \mathrm{SR}^{[n-1]}| \leq \tau \lor n \geq n_{\mathrm{max}} \;$
}{
$\mathrm{SR}^{[n]} \gets \mathrm{SR}^{\ast}$\Comment*[r]{Update previous sum-rate}%
$\mathrm{SR}^{\ast}, \bm{\nu}^{\ast} \gets \mathcal{P}^{\mathrm{final}}_{\mathrm{mix}}\left(\bm{\nu}^{[n]}\right)$\Comment*[r]{Solve convexified subproblem $\mathcal{P}^{\mathrm{final}}_{\mathrm{mix}}$}%
$\bm{\nu}^{[n+1]} \gets \bm{\nu}^{\ast}$\Comment*[r]{Prepare next expansion point}%
$n \gets n + 1$ \\
}(\Comment*[f]{Check for termination})
\Return{$\mathrm{SR}^{\ast}, \bm{\nu}^{\ast}$}
\end{algorithm}

\begin{table}[t]
    \centering
    \caption{Simulation Parameters for Figures 4,5 and 6}
    \label{tab:simulation_Parameters}
    \setlength{\tabcolsep}{5pt}
    \renewcommand{\arraystretch}{1.4}
    \begin{tabular}{|l|c|c|c|c|}
        \hline
        \textbf{Variable} & \textbf{Fig.~4 (a)} & \textbf{Fig.~4 (b)} & \textbf{Fig.~5} & \textbf{Fig.~6} \\
        \hline
        $\tau$ & \multicolumn{4}{c|}{$10^{-5}$} \\
        \hline
        $n_{\mathrm{max}}$ & \multicolumn{4}{c|}{$1000$} \\
        \hline
        Solver & \multicolumn{4}{c|}{MOSEK} \\
        \hline
        $K$ & \multicolumn{4}{c|}{$2$} \\
        \hline        
        $N$ & \multicolumn{4}{c|}{$2$} \\
        \hline
        User 1 position $(x,y)$ & \multicolumn{4}{c|}{$(\qty{0}{\metre}, \qty{100}{\metre})$} \\
        \hline
        User 1 position $(r,\theta)$ & \multicolumn{4}{c|}{$(\qty{100}{\metre}, \qty{0}{\degree})$} \\
        \hline
        User 2 position $(x,y)$ & \multicolumn{3}{c|}{$(\qty{13.92}{\metre}, \qty{99.03}{\metre})$} & Variable \\
        \hline
        User 2 position $(r,\theta)$ & \multicolumn{3}{c|}{$(\qty{100}{\metre}, \qty{8}{\degree})$} & Variable \\
        \hline
        $d$ & \multicolumn{4}{c|}{$\lambda/2$} \\
        \hline
        $f_{\mathrm c}$ & \multicolumn{4}{c|}{$\qty{8}{\giga\hertz}$} \\
        \hline
        $\eta$ & $\qty{10}{\decibel}$ & Variable & \multicolumn{2}{c|}{$\qty{10}{\decibel}$} \\
        \hline
        $\boldsymbol{\rho}^{\mathrm{req}}_{\mathrm{i}}$ & \multicolumn{4}{c|}{$(\qty{0.1}{\bit\per\second\per\hertz}, \qty{0.1}{\bit\per\second\per\hertz})$} \\
        \hline
        $\boldsymbol{\rho}^{\mathrm{req}}$ & \multicolumn{2}{c|}{Variable} & \multicolumn{2}{c|}{$(\qty{0.05}{\bit\per\second\per\hertz}, \qty{0.05}{\bit\per\second\per\hertz})$} \\
        \hline
        $\mathbf{W}^{[0]}_{\mathrm{p}}$ structure & \multicolumn{4}{c|}{MRT, ZF} \\
        \hline
        $\mathbf{W}^{[0]}_{\mathrm{\varsigma}}$ structure & \multicolumn{4}{c|}{ZF, MU-LP-wtc}  \\
        \hline
        $\delta_1$ & \multicolumn{4}{c|}{$[0, 0.1, 0.99]$} \\
        \hline       
        $\delta_2$ & \multicolumn{4}{c|}{$[0.2, 0.5, 0.8]$} \\
        \hline
        $\delta_3$ & \multicolumn{4}{c|}{$[10^{-8}, 0.01, 1.0]$} \\
        \hline
    \end{tabular}
\end{table}

\bibliographystyle{IEEEtran}
\bibliography{IEEEabrv,./bibliography.bib}
\end{document}